\documentclass[fleqn,usenatbib]{mnras}

\usepackage{newtxtext,newtxmath}

\usepackage[T1]{fontenc}

\DeclareRobustCommand{\VAN}[3]{#2}
\let\VANthebibliography\thebibliography
\def\thebibliography{\DeclareRobustCommand{\VAN}[3]{##3}\VANthebibliography}

\usepackage{graphicx}	
\usepackage{amsmath}	
\usepackage{amssymb}	
\usepackage{xspace}
\usepackage{longtable,lscape}
\usepackage{rotating}
\usepackage{multirow,multicol}
\usepackage{balance}

\graphicspath{ {./figures/} }

\newcommand{\msun}{$M_{\odot}$\xspace}
\newcommand{\rsun}{$R_{\odot}$\xspace}
\newcommand{\mearth}{$M_{\oplus}$\xspace}
\newcommand{\rearth}{$R_{\oplus}$\xspace}
\newcommand{\mstar}{\ensuremath{M_{\star}}\xspace}
\newcommand{\rstar}{\ensuremath{R_{\star}}\xspace}

\newcommand{\feh}{\ensuremath{[\mbox{Fe}/\mbox{H}]}\xspace}
\newcommand{\teff}{\ensuremath{T_{\mathrm{eff}}}\xspace}
\newcommand{\logg}{\ensuremath{\log g}\xspace}
\newcommand{\vsini}{\ensuremath{v \sin i_\star}\xspace}
\newcommand{\gcc}{g\,cm$^{-3}$\xspace}

\newcommand{\corot}{\textit{CoRoT}\xspace}
\newcommand{\kepler}{\textit{Kepler}\xspace}
\newcommand{\ktwo}{\textit{K2}\xspace}

\newcommand{\tess}{\textit{TESS}\xspace}

\newcommand{\gaia}{\textit{Gaia}\xspace}
\newcommand{\target}{TOI-763\xspace}

\newcommand{\isochrones}{{\tt isochrones}\xspace}

\newcommand{\vmic}{$V_{\rm mic}$}
\newcommand{\vmac}{$V_{\rm mac}$}

\newcommand{\mgh}{[Mg/H]}

\newcommand{\cah}{[Ca/H]}

\newcommand{\mjup}{$M_\mathrm{J}$}

\newcommand{\halpha}{H$\alpha$}                   

\newcommand{\cai}{Ca\,{\sc I} }

\newcommand{\mgi}{Mg\,{\sc I} }

\newcommand{\kms}{km\,s$^{-1}$}
\newcommand{\ms}{m~s$^{-1}$}

\newcommand{\logrhk}{log\,R$^\prime_\mathrm{HK}$}

\title[TOI-763]{The TOI-763 system: sub-Neptunes orbiting a Sun-like star}

\author[M. Fridlund et al.]{M. Fridlund$^{1,2}$\thanks{E-mail: malcolm.fridlund@chalmers.se},
J. Livingston$^{3}$,
D. Gandolfi$^{4}$,
C.\,M. Persson$^{2}$,
K.\,W.\,F. Lam$^{5}$,
\newauthor
K.\,G. Stassun$^{6}$,
C. Hellier $^{7}$,
J. Korth$^{8}$,
A.\,P. Hatzes$^{9}$,
L. Malavolta$^{10}$,
R. Luque$^{11,12}$,
\newauthor
S. Redfield $^{13}$,
E.\,W. Guenther$^{9}$,
S. Albrecht$^{14}$,
O. Barragan$^{15}$,
S. Benatti$^{16}$,
L. Bouma$^{17}$,
\newauthor
J. Cabrera$^{18}$,
W.D. Cochran$^{19,20}$,
Sz. Csizmadia$^{18}$,
F. Dai$^{21,17}$,
H. J. Deeg$^{11,12}$,
\newauthor
M. Esposito$^{9}$,
I. Georgieva$^{2}$,
S. Grziwa$^{8}$,
L. Gonz\'alez Cuesta$^{11,12}$,
T. Hirano$^{22}$,
\newauthor
J. M. Jenkins$^{23}$,
P. Kabath$^{24}$, 
E. Knudstrup$^{14}$,
D.W. Latham$^{25}$,
S. Mathur$^{11,12}$,
\newauthor
S.\,E.\,Mullally$^{32}$,
N. Narita$^{26,27,28,29,11}$, 
G. Nowak$^{11,12}$,
A.\,O.\,H. Olofsson $^{2}$,
E. Palle$^{11,12}$,
\newauthor
M. P\"atzold$^{8}$,
E. Pompei$^{30}$,
H. Rauer$^{18,5,38}$,
G. Ricker$^{21}$,
F. Rodler$^{30}$,
S. Seager$^{21,31,32}$,
\newauthor
L. M. Serrano$^{4}$,
A. M. S. Smith$^{18}$,
L. Spina$^{33}$,
J.	Subjak$^{34,24}$, 
P. Tenenbaum$^{35}$,
\newauthor
E.B. Ting$^{23}$,
A. Vanderburg$^{39}$,
R. Vanderspek$^{21}$,
V. Van Eylen$^{36}$,
S. Villanueva$^{21}$,
\newauthor
J. N. Winn$^{17}$ 
\\
\\
Authors' affiliations are shown at the end of the manuscript}

\date{Accepted XXX. Received YYY; in original form ZZZ}

\pubyear{2020}

\begin{document}
\label{firstpage}
\pagerange{\pageref{firstpage}--\pageref{lastpage}}
\maketitle
%
\begin{abstract}
We report the discovery of a planetary system orbiting \target (aka CD-39\,7945), a $V=10.2$, high proper motion G-type dwarf star that was photometrically monitored by the \tess\ space mission in Sector 10. We obtain and model the stellar spectrum and find an object slightly smaller than the Sun, and somewhat older, but with a similar metallicity.  Two planet candidates were found in the light curve to be transiting the star. Combining \tess\ transit photometry with HARPS high-precision radial velocity follow-up measurements confirm the planetary nature of these transit signals. We determine masses, radii, and bulk densities of these two planets. A third planet candidate was discovered serendipitously in the radial velocity data. The inner transiting planet, \target\,b, has an orbital period of $P_\mathrm{b}$\,=\,5.6~days, a mass of $M_\mathrm{b}$\,=\,$9.8\pm0.8$\,$M_\oplus$, and a radius of $R_\mathrm{b}$\,=\,$2.37\pm0.10$\,$R_\oplus$. The second transiting planet, \target\,c, has an orbital period of $P_\mathrm{c}$\,=\,12.3~days, a mass of $M_\mathrm{c}$\,=\,$9.3\pm1.0$\,$M_\oplus$, and a radius of $R_\mathrm{c}$\,=\,$2.87\pm0.11$\,$R_\oplus$. We find the outermost planet candidate to orbit the star with a period of $\sim$48~days. If confirmed as a planet it would have a minimum mass of $M_\mathrm{d}$\,=\,$9.5\pm1.6$\,$M_\oplus$. We investigated the \tess\ light curve in order to search for a mono transit by planet~d without success. We discuss the importance and implications of this planetary system in terms of the geometrical arrangements of planets orbiting G-type stars.


\end{abstract}




\section{Introduction}
\label{Intro}

Understanding our origin is a strong driver for science. In astrophysics and space sciences, research in exoplanets and of the Solar system is one way to gain knowledge about our place in the Universe and eventually provide a context for the existence of life on Earth. With the discovery of the first exoplanet in 1995 \citep{MayorQueloz1995}~and the subsequent detection of what is now (July 2020) 4171 confirmed exoplanets\footnote{https://exoplanetarchive.ipac.caltech.edu}, the expansion of this field has led to  new and fantastic discoveries that have changed the pre-1995 predictions of what planetary systems look like. 
 
  In the last $\sim$15 years, a number of space missions (\corot, \citet{Baglin2003, Baglin16}; \kepler, 
 \citet{Borucki2010}; \ktwo, \citet{Howell2014}, and \tess, \citet{Ricker2015}) have been launched with the objective of discovering transiting exoplanets and to derive planetary parameters with high precision.  Together with the physical parameters of the exoplanets, the architecture of the systems (defined as the distribution of different categories of planets within their individual systems) has been of extraordinary interest. Thus one of the most important findings in exoplanetology, so far, is the enormous diversity in the types of planetary systems. While not understood so far, this diversity must reflect the conditions of formation of the systems. In this context, the host star being the dominating body in each system is very important. Among different stellar types, it is especially interesting to study the planetary architectures of G-type host stars since the only known habitable planet, our Earth, orbits such a star. 
 
 There are relatively few planets smaller than Neptune for which both the size and mass has been measured. Only 70 such planets are reported orbiting G-stars in the NASA archive as of June 2020.  Most of these have large uncertainties leading to errors in density of a factor of 2 or more. This is due to the fact that hitherto the exoplanetary space missions have been searching relatively faint stars where although the diameters are known with high precision, the follow-up observations to acquire the planetary masses have usually had large errors. It is therefore important that thanks to the launch of the \tess mission relatively bright stars are now being searched for exoplanets. ESA's future exoplanetary mission PLATO \citep{Rauer2014} will have planets orbiting G-stars as a primary objective when it launches in 2027.

\target is a relatively bright G-type star (Table \ref{tab:stellar}). The detection by \tess of the possible transit of two mini-Neptune planets with radii of 2-3 \rearth, and having  orbital periods of 5.6~days and 12.3~days, respectively, is therefore of significant interest and motivates our detailed study. During the follow-up of \target b and c, deriving masses of 9.8~$M_\oplus$ and 9.3~$M_\oplus$, respectively, we found serendipitously, in the radial velocity data, a signature that could be caused by a third planet of similar mass and orbiting the host star every $\sim$48~days.
Together with a number of recently published systems studied by the \tess mission, \citep*[e.g.][]{Nielsen+2020,Diaz+2020},  \target thus belongs to a still small but growing group of G-type stars hosting a compact planet system. Such a configuration is in sharp contrast to our own solar system, but appears to be quite common among the exoplanetary systems discovered to date \citep{Marcy2014PNAS}.
Studies of such planetary systems promise to lead to a better understanding of their formation process. 

The aim with this paper is to report the characterization of the TOI-763 planet system including investigating the possibility of a third  planet. A secondary objective is to place this system into the proper context. We present the photometry acquired by the \tess spacecraft in Sect.~\ref{sec:tess}. In Sect.~\ref{sec:fup} we detail our follow-up work from the ground, Sect.~\ref{sec:hoststar} presents the derivation of the physical parameters of the host star. In Sect.~\ref{sec:modeling} we describe the modelling and analysis, and derive the parameters of the planets. This is followed by the discussion in Sect.~\ref{sec:disc}. We end the paper with our conclusions in Sect.~\ref{sec:conc}.

\begin{table}
\centering
\caption{Main identifiers, equatorial coordinates, proper motion, parallax, optical and infrared magnitudes, and fundamental parameters of \target.}
\label{tab:stellar}
\begin{tabular}{lrr}
\hline
Parameter & Value & Source \\
\hline
\multicolumn{3}{l}{\it Main identifiers}  \\
\noalign{\smallskip}
\multicolumn{2}{l}  {TIC} {178819686}  & ExoFOP \\
\multicolumn{2}{l}{CD-39 7945}  & CD \\
\multicolumn{2}{l}{2MASS}{J12575245-3945275}  & ExoFOP \\
\multicolumn{2}{l}{UCAC4} {252-056134}  & ExoFOP\\
\multicolumn{2}{l}{WISE} {J125752.37-394528.5}  & ExoFOP\\
\multicolumn{2}{l}{APASS} {18487092}  & ExoFOP\\
\multicolumn{2}{l}{Gaia DR2} {6140553127216043648}  & Simbad\\
\hline
\multicolumn{3}{l}{\it Equatorial coordinates, parallax, and proper motion}  \\
\noalign{\smallskip}
R.A. (J2000.0)	&    12$^\mathrm{h}$57$^\mathrm{m}$52.45$^\mathrm{s}$	& {\it Gaia} DR2 \\
Dec. (J2000.0)	& $-$39$\degr$45$\arcmin$27.71$\arcsec$	                & {\it Gaia} DR2 \\
$\pi$ (mas) 	& $10.4837\pm0.0495$                                    & {\it Gaia} DR2 \\
$\mu_\alpha$ (mas\,yr$^{-1}$) 	& $-76.902 \pm 0.073$		& {\it Gaia} DR2 \\
$\mu_\delta$ (mas\,yr$^{-1}$) 	& $-84.817 \pm 0.055$		& {\it Gaia} DR2 \\
\hline
\multicolumn{3}{l}{\it Optical and near-infrared photometry} \\
\noalign{\smallskip}
$TESS$              & $9.528\pm0.006$     & TIC v8         \\
\noalign{\smallskip}
$G$				 & $9.9992\pm0.0002$	& {\it Gaia} DR2 \\
$B_\mathrm{p}$   & $10.3832\pm0.0005$   & {\it Gaia} DR2 \\
$R_\mathrm{p}$   & $9.4791\pm 0.0012$   & {\it Gaia} DR2 \\
\noalign{\smallskip}
$B$              & $10.855 \pm 0.028$          & APASS \\
$V$              & $10.156 \pm 0.041$          & APASS \\
$g$              & $10.464 \pm 0.034$          & APASS \\
\noalign{\smallskip}
$J$ 			&  $8.858\pm0.029$      & 2MASS \\
$H$				&  $8.554\pm0.023$      & 2MASS \\
$Ks$			&  $8.490\pm0.021$      & 2MASS \\
\noalign{\smallskip}
$W1$			&  $8.422\pm0.023$      & All{\it WISE} \\
$W2$			&  $8.476\pm0.019$      & All{\it WISE} \\
$W3$             & $8.423\pm0.020$      & All{\it WISE} \\
$W4$             & $8.518\pm0.211$      & All{\it WISE} \\
\hline

\end{tabular}
\end{table}
\section{TESS photometry and transit detection}
\label{sec:tess}

\begin{figure}
	\includegraphics[width=\columnwidth]{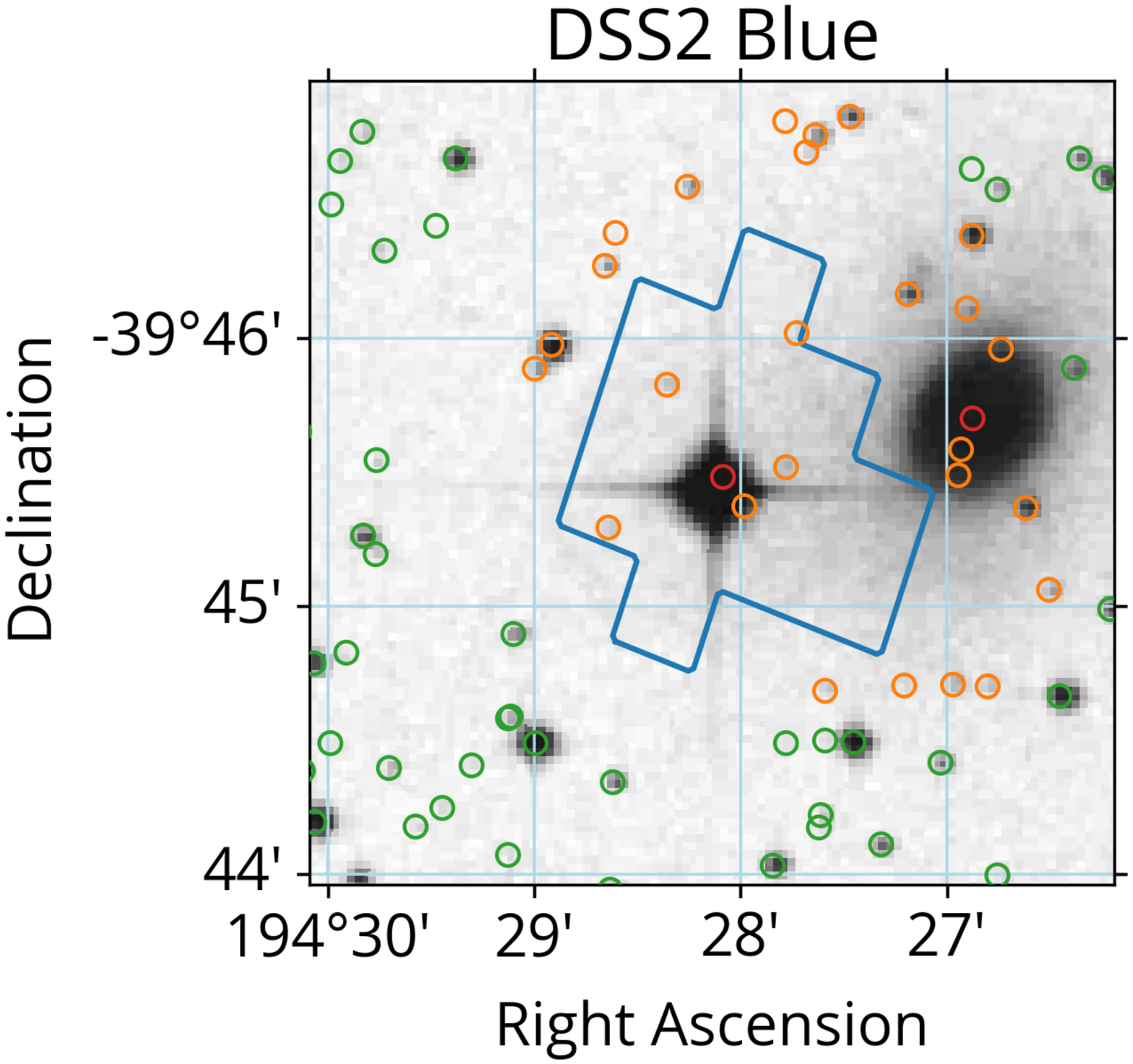}
    \caption{3\arcmin\,$\times$\,3\arcmin\ DSS2 (blue filter) image with the Sectors 10 SPOC photometric aperture outlined in blue. Colored circles denote the positions of \gaia\ DR2 sources within 2\arcmin\ of \target; the red circle inside the aperture is \target (6140553127216043648), the red circle outside the aperture is a galaxy (6140553157278886400), the orange circles are potentially diluting sources, and other sources are in green. We establish a dilution of $<$1\,\% for \target\ based on the flux contributions of all sources.}
    \label{fig:aper}
\end{figure}

\begin{figure*}
	\includegraphics[width=0.9\textwidth]{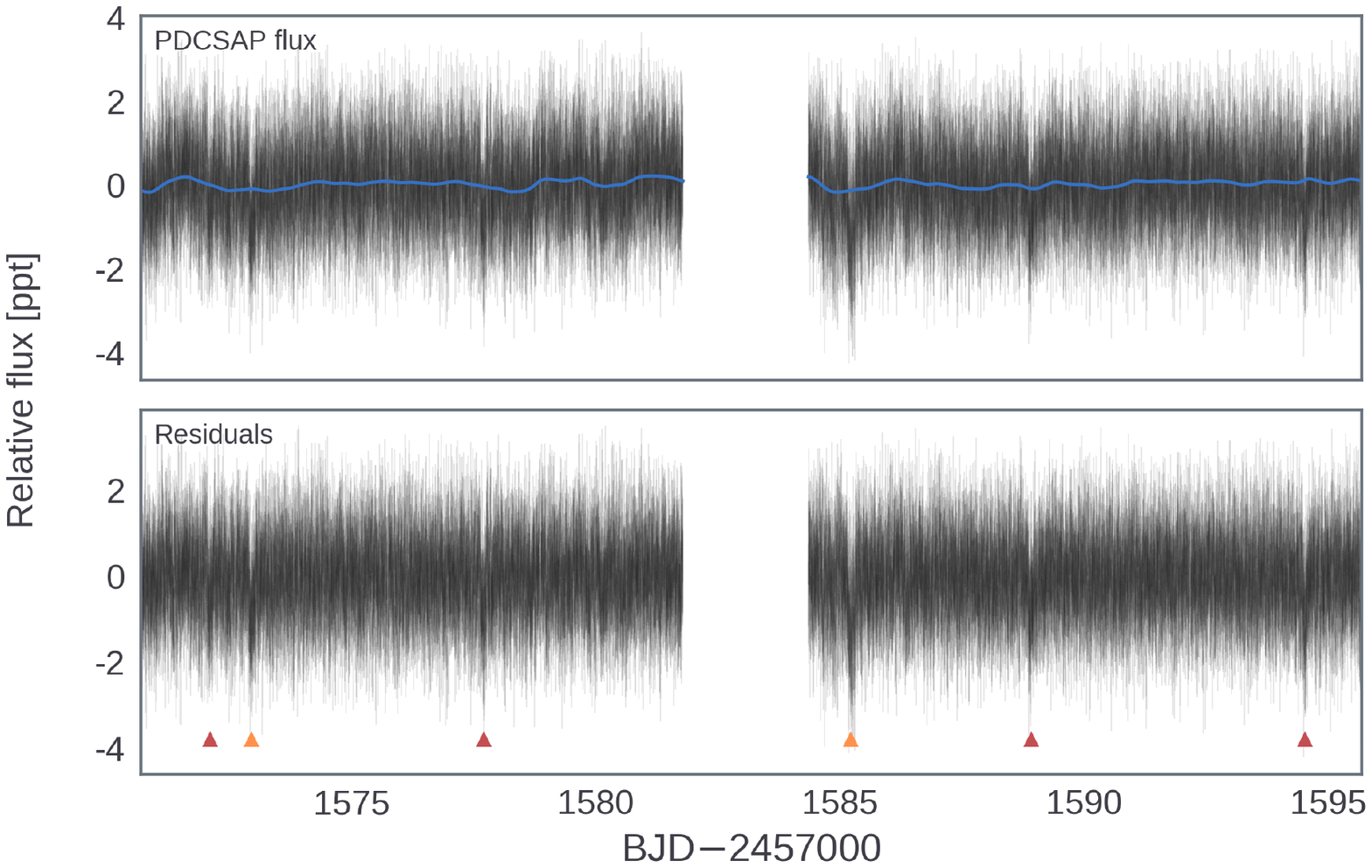}
    \caption{Upper panel: \tess Sector 10 PDCSAP light curve with GP model over-plotted in blue. Lower panel: the ``flattened'' light curve resulting from removing the GP model; the red and orange triangles indicate the transits of planets b and c, respectively.}
    \label{fig:lc}
\end{figure*}

\begin{figure*}
	\includegraphics[width=0.8\textwidth]{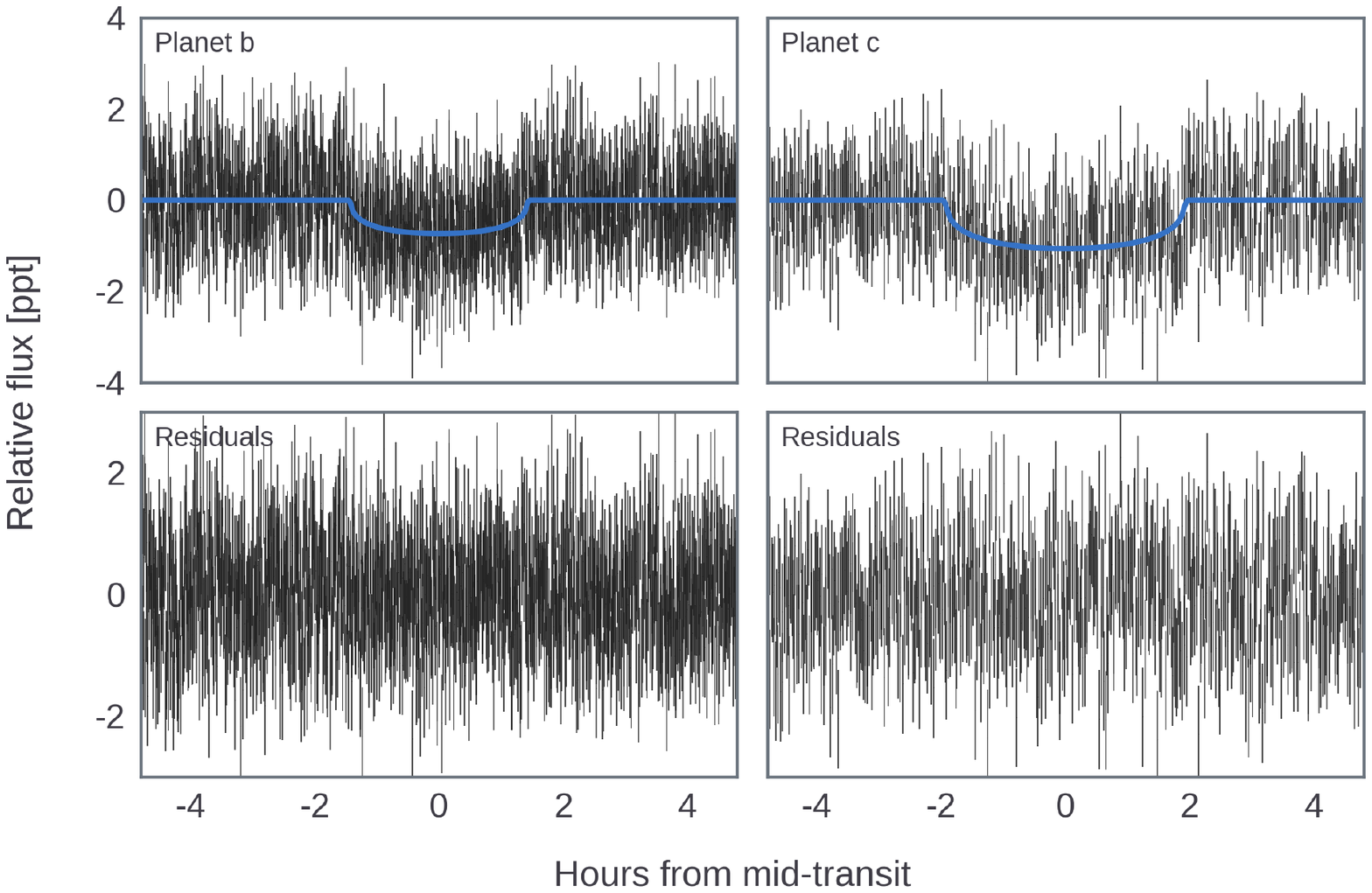}
    \caption{Flattened \tess photometry (black) folded on the best-fitting orbital periods of the two transiting planets, with transit models overplotted in blue (Sect.~\ref{sec:modeling}).}
    \label{fig:transit}
\end{figure*}

\target\ (TIC\,178819686) was observed by \tess\ in Sector 10 between 26 March 2019 and 22 April 2019 (UTC), where the target was imaged on CCD 3 of camera 2. The photometric data were sampled in the 2-min cadence mode and was processed by the science processing operations center \citep[SPOC;][]{Jenkins2016} data reduction pipeline. The SPOC pipeline produces time series light curve using simple aperture photometry (SAP), and the presearch data conditioning (PDCSAP) algorithm was used to remove common instrumental systematics in the light curve \citep{Smith2012,Stumpe2012}. 

We investigated a 3\arcmin$\,\times\,$3\arcmin\  digitized sky survey 2 (DSS-2, blue filter) image centered on \target. Using the Sector 10 SPOC photometric aperture and the positions of \gaia\ data release 2 (DR2) sources we established a dilution of $<$1\,\% for \target\ (Fig.~\ref{fig:aper}). For the light curve and transit analysis of \target, we used the PDCSAP light curve publicly available in the Mikulski Archive for Space Telescopes (MAST)\footnote{\url{https://archive.stsci.edu/tess/}.}. The top panel of Fig.~\ref{fig:lc} shows the PDCSAP light curve of \target.

Transit searches by the SPOC pipeline \citep{Jenkins2002} revealed the presence of two signals at 5.61~d and 12.28~d in the data validation reports \citep{Twicken2018, Li2019}. The detections were announced as planetary candidates via the TOI releases portal\footnote{\url{https://tess.mit.edu/toi-releases/}.}. We iteratively searched the PDCSAP light curve for transit signals using the \emph{d\'etection sp\'ecialis\'ee de transits} (DST) algorithm \citep{Cabrera2012}. The algorithm first applies the Savitzky-Golay method \citep{Savitzky1964,Press2002} to filter variability in the light curve, then uses a parabolic transit model for transit searches. A $12.27\pm0.01$~d transit signal was first detected, which has a transit depth of $813\pm67$~ppm and a duration of $4.07\pm0.18$~h. After filtering the 12.27~d signal, a second transit signal at 5.60~d was detected where transits have a depth of $620\pm54$~ppm and a transit duration of $2.66\pm0.13$~h. Our detection algorithm recovered the transit signals of both TOIs and no further significant periodic transit signal was detected (Fig.~\ref{fig:transit}).

Shallow transits that are close to the noise limit of the light curve may be filtered out by the detection algorithms. We incrementally varied the window size of the Savitzky-Golay filter and visually inspected the light curve of \target\ in order to search for further single transit events. No significant events were found. The Transit Least Square algorithm \citep[\texttt{TLS};][]{Hippke2019} was also implemented to search for single, shallow transit events by fixing the maximum trial period to the observed time baseline of TESS sector 10. This independently confirmed that no single transit event is found above the noise level of the light curve.

\section{Ground-based follow-up observations}
\label{sec:fup}
\subsection{HARPS radial velocity observations}
\label{sec:harps} 

We acquired 74 high-resolution ($R$\,$\approx$\,115\,000) spectra of \target\ using the HARPS fibre-fed Echelle spectrograph \citep[][]{Mayor2003} mounted at the ESO 3.6-m telescope of La Silla observatory (Chile). The observations were performed between 21 June and 01 September 2019 (UTC), as part of our HARPS follow-up program of \tess\ transiting planets (program ID: 1102.C-0923; PI: D. Gandolfi). We used the second fibre of the spectrograph to monitor the sky-background and set the exposure time to 1500\,--\,2400~s depending on sky conditions and schedule constraints. We reduced the HARPS data using the dedicated data reduction software (\texttt{DRS}) and extracted the radial velocity (RV) measurements by cross-correlating the extracted Echelle spectra with a G2 numerical mask \citep{Baranne1996,Pepe2002,Lovis2007}. Following the method described in \citet{Malavolta2017}, we corrected 27 HARPS measurements for scattered moonlight contamination. We also used the \texttt{DRS} to extract the Ca\,{\sc ii} H\,\&\,K lines activity indicator (log\,R$^\prime_\mathrm{HK}$), and three profile diagnostics of the cross-correlation function (CCF), namely, the contrast, the full width at half maximum (FWHM), and the bisector inverse slope (BIS). We finally used the spectrum radial velocity analyser  \citep[\texttt{SERVAL}][]{2018A&A...609A..12Z} to extract four additional activity diagnostics, namely, the chromatic RV index (Crx), differential line width (dLW), and the Na D and H$\alpha$ line indexes.

The \texttt{DRS} HARPS RV measurements and their uncertainties, alongside the barycentric Julian date in barycentric dynamical time (BJD$_\mathrm{TDB}$), the exposure time (T$_\mathrm{exp}$), the signal-to-noise (S/N) ratio per pixel at 5500\,\AA, and the eight activity diagnostics (BIS, FWHM, contrast, dLW, Crx, Na D, H$\alpha$, and \logrhk) are listed in Tables~\ref{table:DRS_RVs} and \ref{table:Serval_Indexes}. 
\subsection{Frequency analysis of the HARPS measurements}
\label{sec:freq}

\begin{figure}
	\includegraphics[width=\columnwidth]{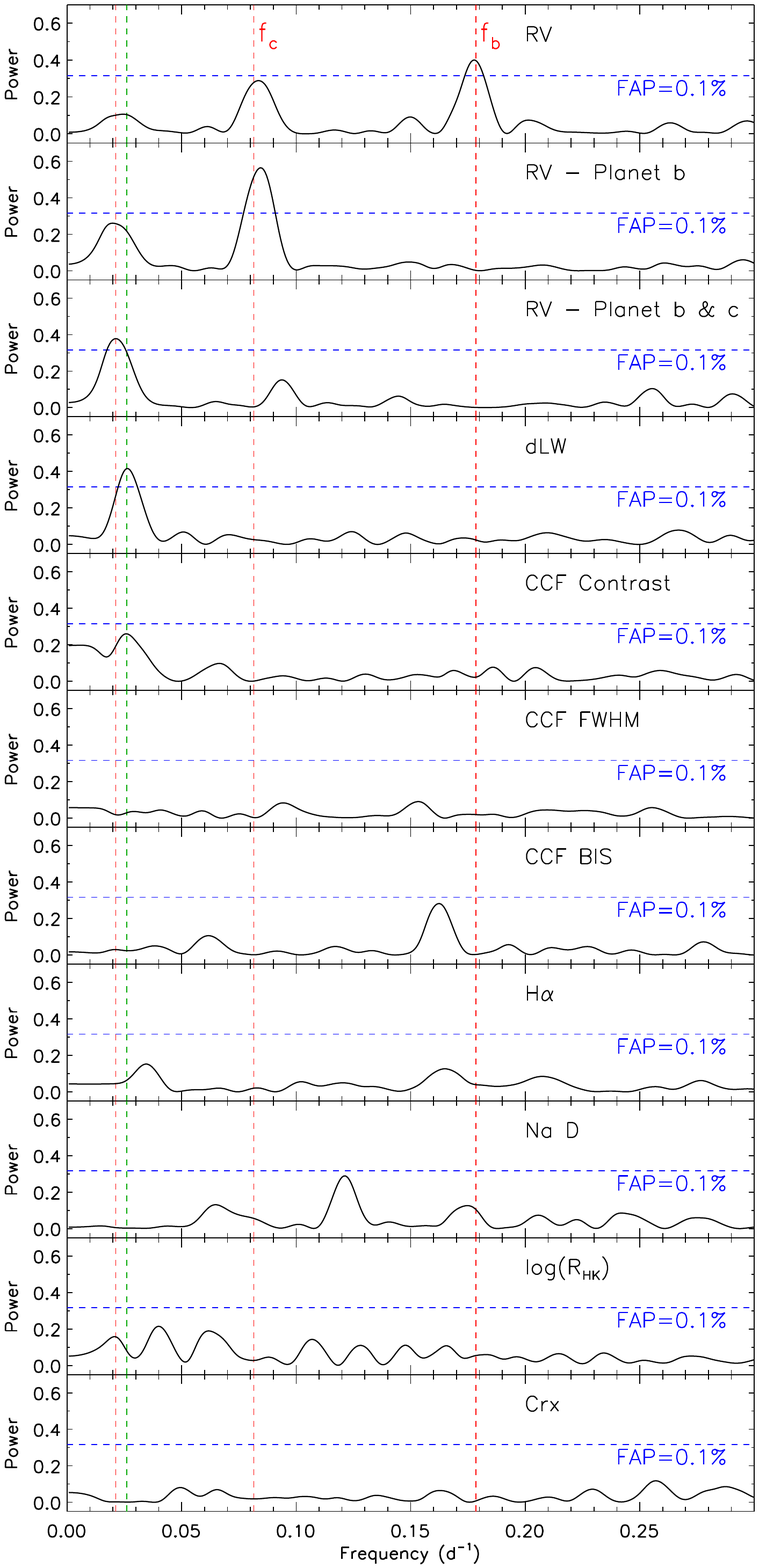}
    \caption{From top to bottom: GLS periodogram of: the HARPS RV measurements (upper panel); the RV residuals following the subtraction of the signal of \target\,b (second panel); the RV residuals following the subtraction of the signals of \target\,b and c (third panel); the dLW (fourth panel); the CCF's contrast (fifth panel); the CCF's FWHM (sixth panel); the CCF's BIS (seventh panel); the H$\alpha$ line index (eighth panel); the Na D lines index (ninth panel); the Ca\,{\sc ii}\,H\,\&\,K lines activity indicator (tenth panel); the Crx index (bottom panel). The dashed vertical red lines mark the orbital frequencies of \target\,b and c (f$_\mathrm{b}$=0.178\,days$^{-1}$ and f$_\mathrm{c}$=0.081\,days$^{-1}$, respectively), and of the additional signal found in the HARPS RVs (0.021\,days$^{-1}$). The dashed green line is the frequency of the dLW peak (fourth panel from the top). The horizontal dashed blue lines mark the false alarm probability of FAP=0.1\,\%, determined using the bootstrap randomization method described in Sect.~\ref{sec:freq}.} 
    \label{fig:GLS_Periodogram}
\end{figure}

We performed a frequency analysis of the \texttt{DRS RV} measurements and \texttt{DRS}/\texttt{SERVAL} activity indicators to look for the Doppler reflex motion induced by the two planets transiting \target\ and unveil the presence of potential additional signals in the time series. 

The generalized Lomb-Scargle (GLS) periodogram \citep{Zechmeister2009} of the HARPS measurements (Fig.~\ref{fig:GLS_Periodogram}, upper panel) shows a significant peak at the orbital frequency of \target\,b (f$_\mathrm{b}$\,=\,0.178\,d$^{-1}$). We derived its false alarm probability (FAP) following the bootstrap randomization method described in \citet{Murdoch+1993}. Briefly, we created 10$^6$ random shuffles of the RV data, while keeping the time stamps fixed, and found that over the frequency range 0--0.50~d$^{-1}$ there were \emph{only} 14 instances where ``random'' data had power in the periodogram greater than the peak seen at f$_\mathrm{b}$. The FAP is thus 14/10$^6$\,=\,1.4\,$\times$\,10$^{-5}$.

The GLS periodogram of the HARPS RV residuals (Fig.~\ref{fig:GLS_Periodogram}, second panel) -- following the subtraction of the Doppler signal induced by the inner planet -- displays a significant peak almost at the orbital frequency of \target\,c (f$_\mathrm{c}$\,=\,0.081\,d$^{-1}$). We applied the same procedure to estimate its FAP and found that none of the periodograms computed from the 10$^6$ random shuffles of the RV residuals has a power greater than the peak at f$_\mathrm{c}$. The FAP is thus $<$\,10$^{-6}$ over the frequency range 0--0.50\,d$^{-1}$.

The two Doppler signals at f$_\mathrm{b}$ and f$_\mathrm{c}$ have no counterpart in the periodograms of the eight activity indicators (Fig.~\ref{fig:GLS_Periodogram}), confirming the planetary nature of the two transit signals detected in the \tess\ light curve.

Following the subtraction of the Doppler reflex motion induced by the two transiting planets, the periodogram of the HARPS RV residuals shows an additional peak at about 48~d (0.021\,d$^{-1}$), whose FAP is equal to 5.1$\times$\,10$^{-5}$ (Fig.~\ref{fig:GLS_Periodogram}, third panel, leftmost red dashed line). This signal has no counterpart in the periodograms of the BIS, FWHM, Crx, \logrhk, H$\alpha$, and Na D lines. However, we found that the periodogram of the dLW shows a significant peak at 0.026\,d$^{-1}$, corresponding to a period of about 38.4 days (green dashed line in Fig.~\ref{fig:GLS_Periodogram}). The CCF contrast shows also a peak at 0.026\,d$^{-1}$, although it is not significant (FAP\,$\approx$\,0.1\,\%). The difference between the two frequencies (0.005\,d$^{-1}$) is about three times smaller than the spectral resolution\footnote{The spectral resolution is defined as the inverse of the baseline. For our HARPS follow-up the baseline is $\sim$73 d, which translates into a resolution of about 0.014\,d$^{-1}$.} of our RV time-series (0.014\,d$^{-1}$). This implies that the two peaks at 38.4 and 48~d remain unresolved in our data-set and we cannot assess whether they arise from the same source or not. If the two peaks have a common origin, then they are likely associated to the presence of active regions carried around by stellar rotation. Alternatively, the peak at 48~d might be due the presence of an additional outer planet, whereas the peak detected in the periodogram of the dLW might be associated to magnetic activity coupled with stellar rotation.
\subsection{Ground based Photometry}
\label{sec:WASP}

WASP-South, the southern station of the WASP project \citep{Pollaco2006}, consists of an array of 8 cameras.  From 2006 to 2012 the cameras used  200-mm, f/1.8 lenses with a filter spanning 400--700 nm. From 2012 to 2016 they used 85-mm, f/1.2 lenses with an SDSS-$r$ filter \citep{AMSSmith+2014}. On clear nights, available fields were rastered with a typical 10-min cadence.  WASP-South observed the field of TOI-763 over typically 120 nights per year, accumulating 24\,000 data points with the 200-mm lenses and then 43\,000 datapoints with the 85-mm lenses.   We searched the accumulated data on TOI-763 for a rotational modulation, using the methods presented in \citet{2011PASP..123..547M}, but find no significant periodicities. For periods from 2~days up to $\sim$\,100~days we find an upper limit of 1~mmag for any rotational modulation.  

\section{Host star fundamental parameters}
\label{sec:hoststar} 

\subsection{Analysis of the optical spectrum}
\label{sec:SME}

The fundamental parameters of the host star are important for deriving precise values for the planetary masses, radii and thus bulk densities. Most important in this analysis is the effective temperature of the star, \teff, and, lacking an interferometrically determined diameter of \target, we derived \teff\ from the optical HARPS data. Co-adding the 74 individual high resolution HARPS spectra resulted in a very high signal-to-noise spectrum (about 300 per pixel at 5500\AA). We then compared the co-added HARPS spectrum with modeled synthetic spectra. For this we used the Spectroscopy Made Easy (SME) package \citep{Valenti1996,Piskunov2017} version 5.22, with atomic parameters from the VALD database \citep{Piskunov95}. SME calculates synthetic spectra based on a number of stellar parameters using a grid of stellar atmospheric models. The grid we used in this case is based on the Atlas-12 models \citep{Kurucz2013}. The calculated spectrum was then compared to the observed spectrum and an iterative $\chi^2$ minimization procedure was followed until no improvement was achieved. We refer to recent papers, e.g., \citet{Persson2018} and \citet{Malcolm2017} for details about the method. In order to limit the number of free parameters we used empirical calibrations for the \vmic\ and \vmac\ turbulence velocities \citep{Bruntt2010, Doyle2014}. The value of \teff\ was determined from fitting the Balmer \halpha~line wings. We then used the derived \teff\ to fit a large sample of Fe {\sc i}, \mgi and \cai lines, all with well established atomic parameters in order to derive the abundance, \feh, and the \logg. We found the star to be slowly rotating, \vsini~=\,1.7~$\pm$\, 0.4 \kms. This is consistent with the low activity as detailed in Section~\ref{sec:freq} and \ref{sec:WASP}. The star is somewhat cooler than our Sun, with an effective temperature as derived from the \halpha~line wings of \teff\,=\,$5450 \pm 60$~K. 
Using this value for \teff\ we found the \feh~to be $0.01\,\pm\,0.05$ and the surface gravity \logg~to be $4.45\,\pm\,0.05$ (Table~\ref{tab:star_results}).

As a sanity check we also analysed the same co-added spectrum using the package SpecMatch-Emp \citep{Yee2017}. This is a public software that matches a large part of the spectrum to a library of stellar spectra with well-established fundamental parameters. We refer to \citet{Hirano2018} to describe the special procedure we used in order to match our data to the format used as input in the SpecMatch-Emp code. The library used in this code was created, using stars that are either eclipsing binaries or that have radii determined through interferometry. We obtained a stellar radius of $R_\star\,=\,1.126\,\pm\,0.18$~\rsun, an effective temperature of \teff=\,$5444\pm110$~K, and an iron abundance of \feh$\,=\,-0.09\,\pm\,0.09$. The latter two values are in agreement with the results from the SME analysis. Because of the higher precision in the SME analysis, the final adopted value of \teff\ for \target is \mbox{$5450\,\pm\,60$~K}. The error is the internal errors in the synthesising of the spectra and does not include the inherent errors of the model grid itself, as well as those errors caused by using 1-D models. Finally, \target is in the \tess-Gaia catalogue of \citet{Carillo+2020} where the DR2 Gaia astrometry is used to compute the membership probabilities in the galactic thin disk, thick disk and halo to be 0.95490, 0.04509 and 0.00001 respectively, consistent with the solar like metallicity derived from the spectral analysis.

\begin{table}
\centering
\caption{Stellar parameters of the \target system. Values from spectral synthesis (SME) and the analysis of the SED. 
\label{tab:star_results}}
\begin{tabular}{lllll}
\hline
Parameter & Unit & \textit{Star} &Source & \\
\hline
\noalign{\smallskip}
\teff 	&(K)& $5450\pm60$  & \ref{sec:SME}&\\
\logg 	&(cgs)	& $4.45 \pm 0.050$ & \ref{sec:SME} &\\
\feh  	& (dex)	& $0.01 \pm 0.05$  & \ref{sec:SME} &\\
\cah  	& (dex)	& $0.05 \pm 0.05$  & \ref{sec:SME} &\\
\mgh  	& (dex)	& $0.12 \pm 0.05$  & \ref{sec:SME} &\\
$A_\mathrm{V}$ & mag & $0.02 \pm 0.02$ & \ref{sec:hostrad}&\\
$P_{\rm rot}/\sin i$ &(days) & $27 \pm 16$ & \ref{sec:hostrot}& \\
\hline
\end{tabular}
\end{table}

\subsection{Stellar radius via spectral energy distribution}
\label{sec:hostrad}
\begin{figure}
    \centering
    \includegraphics[width=6.3cm,angle=90,trim=70 90 90 95,clip]{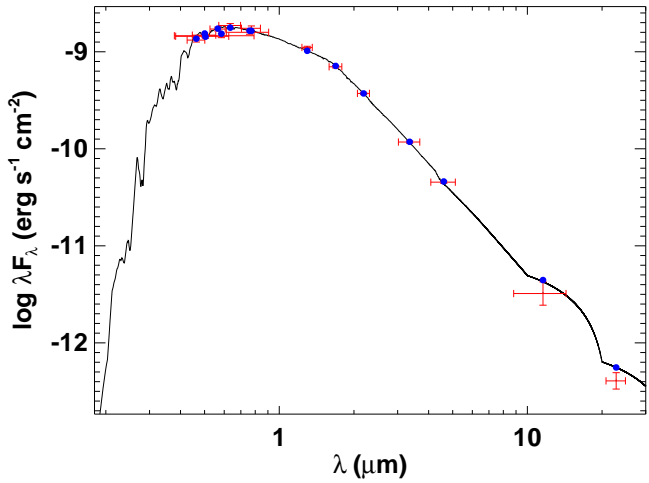}
    \caption{Spectral energy distribution. Red symbols represent the observed photometric measurements, where the horizontal bars represent the effective width of the passband. Blue symbols are the model fluxes from the best-fit Kurucz atmosphere model (black). 
\label{fig:sed}}
\end{figure}

We performed an analysis of the broadband Spectral Energy Distribution (SED) together with the {\it Gaia\/} DR2 parallax in order to determine an empirical measurement of the stellar radius, following the procedures described in \citet{Stassun:2016}, \citet{Stassun:2017}, and \citet{Stassun:2018}. We pulled the $BVgri$ magnitudes from APASS, the $JHK_S$ magnitudes from {\it 2MASS}, the W1--W4 magnitudes from {\it WISE}, and 
the $G, G_{\rm BP}$~and $G_{\rm RP}$ magnitudes from {\it Gaia} (Table~\ref{tab:stellar}). Together, the available photometry spans the full stellar SED over the wavelength range 0.4\,--\,22~$\mu$m (Fig.~\ref{fig:sed}). 

We performed a fit using Kurucz stellar atmosphere models, with the priors on \teff, \logg, and \feh from the spectroscopic analysis (Table~\ref{tab:star_results}). The remaining free parameter is the extinction ($A_V$), which we limited to the maximum permitted for the star's line of sight from the \citet{Schlegel:1998} dust maps. The resulting fit is very good (Fig.~\ref{fig:sed}) with a reduced $\chi^2$ of 1.4 and $A_V = 0.02\,\pm\,0.02$. Integrating the model SED gives the bolometric flux at Earth of \mbox{$F_{\rm bol} = 2.345 \pm 0.055 \times 10^{-9}$ erg~s$^{-1}$~cm$^{-2}$}. Taking the $F_{\rm bol}$ and \teff\ together with the {\it Gaia\/} parallax, adjusted by $+0.08$~mas to account for the systematic offset reported by \citet{StassunTorres:2018}, gives the stellar radius as \mbox{\rstar $= 0.910\,\pm\,0.020$~\rsun} (See Table \ref{tab:thehatefuleight}). 

\subsection{Stellar mass via radius and surface gravity}
\label{sec:hostmass}
The empirical stellar radius determined above affords an opportunity to estimate the stellar mass empirically as well, via the spectroscopically determined surface gravity (\mbox{\logg $ = 4.50\,\pm\,0.05$}), which gives \mstar $= 0.95\,\pm\,0.12$~\msun. This is consistent with that estimated via the eclipsing-binary based relations of \citet{Torres+2010}, which gives 
\mbox{\mstar $=0.95\,\pm\,0.06$~\msun} 
(See Table \ref{tab:thehatefuleight}). 

\subsection{Stellar age via activity and rotation}
\label{sec:hostrot}

\begin{figure}
    \centering
    \includegraphics[width=8.5cm,trim=100 65 60 80,clip]{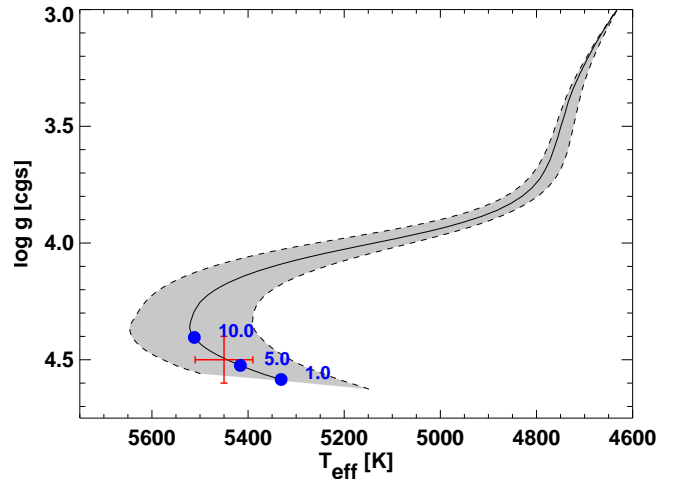}
    \caption{H-R diagram representing the observed \teff\ and \logg\ (red symbol) in relation to a stellar evolution model from the Yonsei-Yale grid for the star's inferred mass and observed \feh; the gray swathe corresponds to the uncertainty in the inferred mass. Model ages in Gyr are represented as blue symbols.}
    \label{fig:hrd}
\end{figure}

We used our HARPS observations to estimate the stellar age from its chromospheric activity, as measured by the Ca $R'_{\rm HK}$ index, which we determined to be $-4.98\pm0.03$. Using the activity-age relations of \citet{MamajekHillenbrand:2008}, we obtained from $R'_{\rm HK}$ and the star's $B-V$ color, an age of $\tau = 6.2 \pm 0.6$~Gyr. This is consistent with the age implied by the star's position in the H-R diagram in comparison to the Yonsei-Yale stellar evolution models (Fig.~\ref{fig:hrd}). 

The \citet{MamajekHillenbrand:2008} relations also give a predicted rotation period for the star, based again on the $R'_{\rm HK}$ activity and $B-V$ color. The derived rotation period of \mbox{$P_{\rm rot} = 32.1\,\pm\,1.3$~days} is consistent with value inferred from the stellar radius and the spectroscopic \vsini, which gives $P_{\rm rot}/\sin i = 27 \pm 16$~days. Moreover, the GLS periodogram of the differential line width activity indicator shows a peak at $\sim$38.4\,d (FAP\,<\,0.1\,\%), which could be associated to stellar rotation, in agreement with the previous estimates. Altogether, all of the evidence is consistent with a slowly rotating star that is a bit older than the Sun.

\subsection{Stellar parameters from \isochrones}
\label{sec:isochrones}
We used the Python {\tt isochrones} \citep{2015ascl.soft03010M} interface to the MIST stellar evolution models \citep{2016ApJ...823..102C} to infer a uniform set of stellar parameters. We performed a fit to 2MASS $JHK$ photometry \citep{2006AJ....131.1163S} and {\it Gaia} DR2 parallax \citep{2016A&A...595A...1G, 2018A&A...616A...1G} using {\sc MultiNest} \citep{2013arXiv1306.2144F} to sample the joint posterior. We placed priors on \teff and \feh based on the spectroscopic results from SME, using a \teff uncertainty of 100~K to account for systematic errors. We obtained the following parameter estimates: \teff= $5571\,\pm\,68$~K, \logg= $4.505^{+0.022}_{-0.025}$, \feh= $0.01 \pm 0.05$~dex, \mstar= $0.936 \pm 0.031$~$\mathrm{M_\odot}$, \rstar= $0.896\,\pm\,0.013$~$\mathrm{R_\odot}$, age= $3.8^{+2.8}_{-2.3}$~Gyr, distance= $95.48 \pm 0.98$~pc, and $A\mathrm{_V}$=$0.08^{+0.10}_{-0.05}$~mag, in good agreement with the values above.
\begin{table}
\scriptsize
\caption{Stellar mass and radius of \target~as derived from different methods.} 
\label{tab:thehatefuleight}
\begin{tabular}{llll}
\hline
 Origin &\rstar(\rsun) &\mstar(\msun) & Note \\
\hline
ExoFOP & $0.9068 \pm 0.047$ & $0.97 \pm 0.1146$& TICv8 \\ 
Specmatch & $1.126 \pm 0.18$& - & \ref{sec:SME} \\
SED & $0.910 \pm 0.02$& $0.95 \pm 0.12$ & \ref{sec:hostrad}, \ref{sec:hostmass}\\
Isochrones$^1$ & $0.896 \pm 0.013$ & $0.936 \pm 0.031$ & \ref{sec:isochrones}\\
RV/transit & $0.899 \pm 0.013$ & $0.915 \pm 0.028$ & \ref{sec:joint}\\
Num. Model$^2$ & $0.96 \pm 0.031$& $0.96 \pm 0.07$ & \ref{sec:hostrad}, \ref{sec:hostmass}\\
\hline
$^1$ Adopted for the modelling\\
$^12$ According to \citet{Torres+2010}

\end{tabular}
\end{table}
 \section{RV and transit analysis}
\label{sec:modeling}

\subsection{Preliminary RV analysis}
\label{sec:rv}

Using the results of our frequency analysis we fit the HARPS RV data using \texttt{RadVel}\footnote{\url{https://github.com/California-Planet-Search/radvel}} \citep{2018PASP..130d4504F}, enabling us to perform RV model selection and estimate system parameters. We tested eight different models: two circular orbits (``2c''); two eccentric orbits (``2e''); three circular orbits (``3c''); three eccentric orbits (``3e''); two circular orbits with a Gaussian Process (GP) noise model (``2cGP''); two eccentric orbits with a GP noise model (``2eGP''); two eccentric orbits (inner planets) and one circular orbit (outer planet); one circular orbit (inner planet) and one eccentric orbit (outer planet). We used a quasi-periodic GP kernel \citep[e.g.][]{2014MNRAS.443.2517H,2015ApJ...808..127G,2017AJ....154..226D} with a Gaussian prior on the period hyper-parameter based on the stellar rotation period estimated in Section~\ref{sec:hoststar},  $P_{\rm rot} = 27 \pm 16$~days and assuming zero obliquity. We present the model comparison in Table~\ref{tab:models}, including both the commonly used Bayesian Information Criterion (BIC) and the Akaike Information Criterion (AICc; corrected for small sample sizes). The ``3c'' model (3 circular orbits) is strongly favored over the other models by both the BIC and the AIC, suggesting that eccentricity in the system is low. The MCMC parameter estimates from this model are presented in Table~\ref{tab:rv}.

\begin{table}
\scriptsize
\caption{RV model comparison (see Section~\ref{sec:rv}). \label{tab:models}}
\begin{tabular}{lccccccr}
\hline
Model &     AICc & BIC & $N_\mathrm{free}$ & $N_\mathrm{data}$ &    RMS$^a$ & $\ln{\mathcal{L}}^b$ \\
\hline
3c &   315.06 &   336.14 &     11 &     74 &   1.70 &  -135.96 \\
1c2e &   322.28 &   348.56 &     15 &     74 &   1.62 &  -133.55 \\
2e1c &   323.59 &   349.87 &     15 &     74 &   1.64 &  -134.25 \\
3e &   327.31 &   355.55 &     17 &     74 &   1.60 &  -132.78 \\
2cGP &   334.28 &   356.81 &     12 &     74 &   1.69 &  -147.14 \\
2c &   341.21 &   357.43 &      8 &     74 &   2.15 &  -143.54 \\
2eGP &   342.91 &   370.23 &     16 &     74 &   1.62 &  -145.24 \\
2e &   346.73 &   369.26 &      12 &     74 &   2.04 &  -140.85 \\
\hline
\multicolumn{7}{l}{$^a$ Root mean square of the data minus the model.}\\
\multicolumn{7}{l}{$^b$ Log-likelihood of the data given the model.}\\
\end{tabular}
\end{table}
\begin{table}
\scriptsize
\caption{MCMC posteriors from the ``3c'' RV model (see Sect.~\ref{sec:rv}). \label{tab:rv}}
\begin{tabular}{lrrr}
\hline
Parameter & Cred. Interval & Max. Likelihood & Units \\
\hline
{\it Free} & & &  \\
\noalign{\smallskip}
  $P_{b}$ & $5.60501\pm 0.00094$ & $5.605$ & days \\
  $T\rm{conj}_{b}$ & $1572.1029^{+0.0029}_{-0.003}$ & $1572.1029$ & BTJD \\
  $K_{b}$ & $3.79\pm 0.31$ & $3.79$ & m s$^{-1}$ \\
  $P_{c}$ & $12.2752^{+0.004}_{-0.0038}$ & $12.2752$ & days \\
  $T\rm{conj}_{c}$ & $1572.9688^{+0.0032}_{-0.0031}$ & $1572.9687$ & BTJD \\
  $K_{c}$ & $2.78\pm 0.31$ & $2.77$ & m s$^{-1}$ \\
  $P_{d}$ & $47.7^{+2.5}_{-1.2}$ & $47.7$ & days \\
  $T\rm{conj}_{d}$ & $1545.6^{+4.0}_{-7.2}$ & $1545.6$ & BTJD \\
  $K_{d}$ & $1.82^{+0.31}_{-0.32}$ & $1.84$ & m s$^{-1}$ \\
  $\gamma_{\rm HARPS}$ & $-0.37\pm 0.21$ & $-0.37$ &  \\
  $\sigma_{\rm HARPS}$ & $1.18^{+0.24}_{-0.23}$ & $1.06$ &  \\
  \hline
{\it Derived} & & &  \\
\noalign{\smallskip}
  $M_b\sin i$ & $10.02^{+0.86}_{-0.85}$ & $10.37$ & \mearth \\
  $M_c\sin i$ & $9.5\pm 1.1$ & $9.5$ & \mearth \\
  $M_d\sin i$ & $9.8\pm 1.7$ & $8.1$ & \mearth \\
\hline
\end{tabular}
\end{table}

\subsection{Joint RV and transit analysis}
\label{sec:joint}

We jointly fit the HARPS RVs and \tess light curve using {\tt exoplanet}\footnote{\url{https://docs.exoplanet.codes/en/stable/}.} \citep{exoplanet}, {\tt Starry} \citep{starry}\footnote{\url{https://rodluger.github.io/starry/v1.0.0/}.}, and {\tt PyMC3}\footnote{\url{https://docs.pymc.io/}.} \citep{pymc3}. We first used a GP model with a Mat\'ern-3/2 kernel to fit the out-of-transit variability in the \tess light curve
(Fig.~\ref{fig:lc}). To achieve this in an accurate and efficient manner, we first masked out the transits and then binned the data by a factor of 100 ($\sim$3.3 hour bins). We then conducted a joint fit to the RVs and the ``flattened'' light curve resulting from the removal of the best-fit GP signal, including mean flux ($\langle f \rangle$) and white noise ($\sigma_{TESS}$) parameters for the photometry. For efficient sampling we used quadratic limb darkening coefficients ($u_1$, $u_2$) under the transformation of \citet{exoplanet:kipping13}. We used minimally informative priors for all parameters except for the host star mass and radius, which were Gaussian and based on our results in Section~\ref{sec:hoststar}. To reduce the possibility of underestimated uncertainties, we allowed the eccentricity of the two transiting planets to float, and included jitter ($\sigma_\mathrm{HARPS}$) and mean velocity ($\gamma_\mathrm{HARPS}$) parameters for the RV data.

\begin{figure*}
	\includegraphics[width=0.8\textwidth]{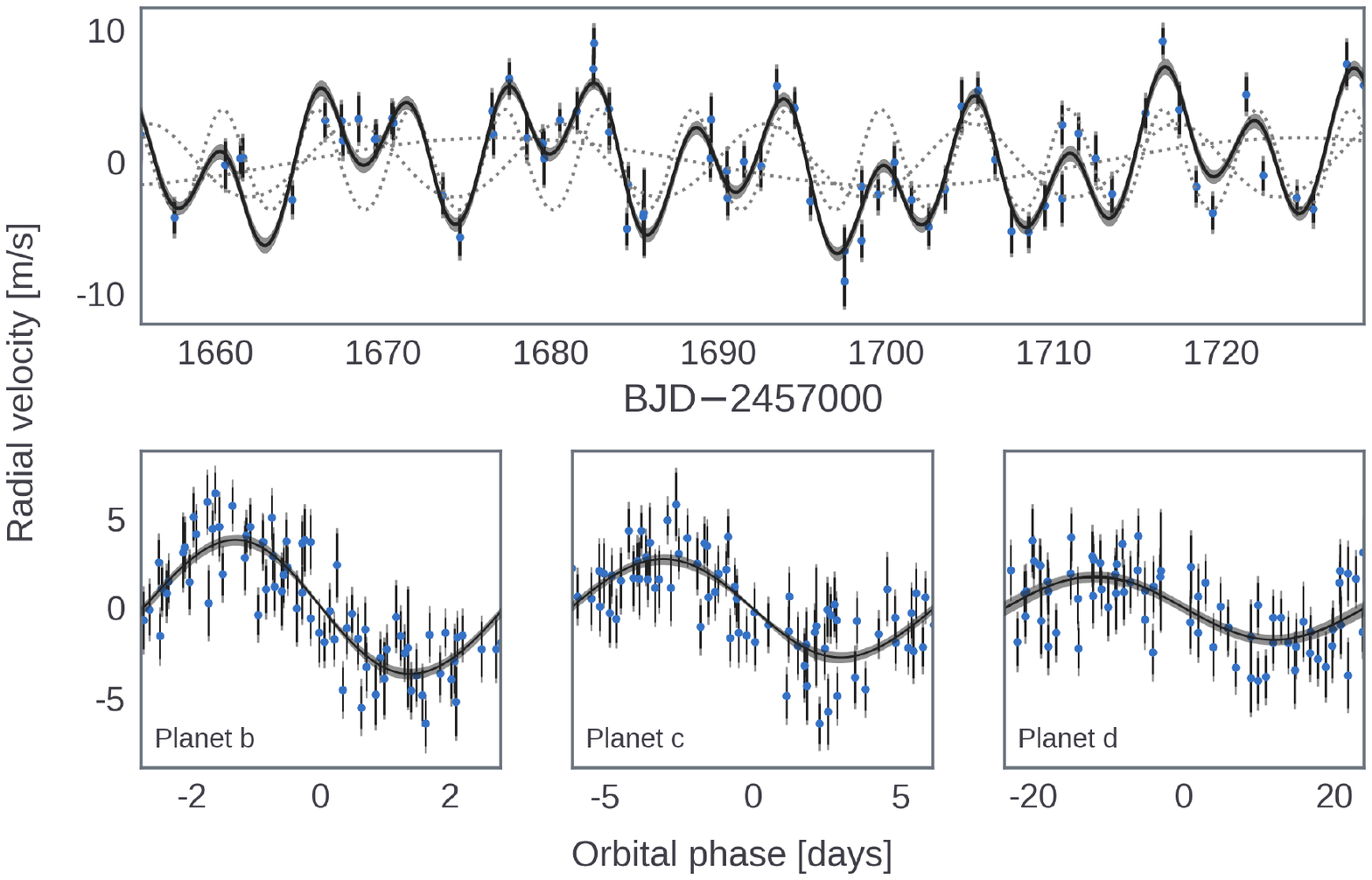}
    \caption{HARPS data (blue) and posterior constraints with 1~$\sigma$ credible region (black). Upper panel: full orbital solution with individual components shown as dotted lines. Lower panel: HARPS data folded on the orbital period of each planet, after subtracting the signals of the other planets. The measured error bars are in black, and the error bars taking into account the jitter ($\sigma_\mathrm{HARPS}$) are in gray.}
    \label{fig:rv2}
\end{figure*}

We used the gradient-based {\tt BFGS} algorithm \citep{NoceWrig06} implemented in {\tt scipy.optimize} to find initial maximum a posteriori (MAP) parameter estimates. We used these estimates to initialize an exploration of parameter space via ``no U-turn sampling'' \citep[NUTS,][]{Hoffman:Gelman:2014}, an efficient gradient-based Hamiltonian Monte Carlo (HMC) sampler implemented in {\tt PyMC3}. After sampling, the Gelman-Rubin statistic \citep{Gelman:Rubin:1992} was $<$1.001 and the sampling error was $\lesssim$1\% for all parameters, indicating the sampler was well-mixed and yielded a sufficient number of independent samples. We present the resulting parameter estimates in Table~\ref{tab:results2}, and the data and posterior constraints from the model in Fig.~\ref{fig:rv2}. There are no significant changes in the derived semi-amplitudes of the planets from using DRS or \texttt{serval} extracted RVs. 

\begin{table*}
\centering
\caption{DRS Joint RV and transit modeling results. Note that the values for the planet candidate d  are tentative. \label{tab:results2}}
\begin{tabular}{lcccc}
\hline
Parameter & Unit & \textit{Star} & & \\
\noalign{\smallskip}
$M_\star$ & $M_\odot$ & $0.917 \pm 0.028$ \\
$R_\star$ & $R_\odot$ & $0.897 \pm 0.013$ \\
$u_1$ & $\cdots$ & $0.78 \pm 0.37$ \\
$u_2$ & $\cdots$ & $-0.08^{+0.43}_{-0.34}$ \\
$\langle f \rangle$ & ppm & $-17 \pm 13$ \\
$\sigma_\mathrm{TESS}$ & ppm & $811 \pm 9$ \\
$\gamma_\mathrm{HARPS}$ & $\mathrm{m}\,\mathrm{s}^{-1}$ & $-0.38 \pm 0.21$ \\
$\sigma_\mathrm{HARPS}$ & $\mathrm{m}\,\mathrm{s}^{-1}$ & $0.13^{+0.18}_{-0.21}$ \\
\hline
Parameter & Unit & \textit{Planet b} & \textit{Planet c} & \textit{Planet candidate d} \\
\noalign{\smallskip}
$T_0$ & BTJD & $1572.1020 \pm 0.0031$ & $1572.9661^{+0.0071}_{-0.0044}$ & $1593.1680 \pm 5.4930$ \\
$P$ & days & $5.6057 \pm 0.0013$ & $12.2737^{+0.0053}_{-0.0077}$ & $47.7991 \pm 2.7399$ \\
$b$ & $\cdots$ & $0.51 \pm 0.35$ & $0.51 \pm 0.35$ & $\cdots$ \\
$e$ & $\cdots$ & $0.04^{+0.04}_{-0.03}$ & $0.04^{+0.04}_{-0.03}$ & $\equiv 0$ \\
$\omega$ & deg. & $42^{+57}_{-84}$ & $62^{+78}_{-153}$ & $\equiv 0$ \\
$R_P$ & $R_\star$ & $0.023 \pm 0.001$ & $0.027 \pm 0.001$ & $\cdots$ \\
$R_P$ & $R_\oplus$ & $2.28 \pm 0.11$ & $2.63 \pm 0.12$ & $\cdots$ \\
$M_P$ & $M_\oplus$ & $9.79 \pm 0.78$ & $9.32 \pm 1.02$ & $9.54 \pm 1.59^1$ \\
$\rho_P$ & g\,cm$^{-3}$ & $4.51^{+0.83}_{-0.66}$ & $2.82^{+0.54}_{-0.47}$ & $\cdots$ \\
$a$ & au & $0.0600 \pm 0.0006$ & $0.1011 \pm 0.0010$ & $0.2504^{+0.0093}_{-0.0105}$ \\
$T_\mathrm{eq}$ & K & $1038 \pm 16$ & $800 \pm 12$ & $509 \pm 12$ \\
\hline
\multicolumn{5}{l}{$^1$ $m \sin i$} \\
\hline
\end{tabular}
\end{table*}

As a sanity check, we also performed a joint analysis of the \tess\ and HARPS time series using \texttt{pyaneti} \citep{Barragan2019}. The \texttt{pyaneti} code utilizes Bayesian approaches coupled with Markov chain Monte Carlo sampling to perform multi-planet radial velocity and transit data fitting. We fitted the HARPS RVs using two Keplerians for the two transiting planets discovered by \tess, and one sine-curve for the additional Doppler signal found in the HARPS RVs. We modelled the transiting light curves using the limb-darkened quadratic model by \citet{MandelAgol}. We adopted uniform priors for all the fitted parameters but the limb darkening coefficients, for which we used Gaussian priors based on \citet{Claret2017}'s \tess\ coefficients. The results agree well with those previously obtained. In particular, the masses and radii agree within 1~$\sigma$, or less, indicating the parameter estimates are robust.

The stellar spectroscopic parameters are consistent with a very low level of activity.  We also detect no rotational modulation in either the HARPS activity indicators (Sect.~\ref{sec:freq}), or in the WASP light curve (Sect.~\ref{sec:WASP}). Given this, we conclude that \target\ was about as active as our own Sun is in the quiet part of the 11-year solar cycle at the time our observations were carried out. This is also consistent with the projected rotational velocity of \vsini=\,1.7~\kms, thus indicating a mature G-type star (Sect.~\ref{sec:isochrones}). Together with the indications of a rotation period of around 30d from the v sin i and \halpha
~index, these circumstances make it highly unlikely that the modulation found in the HARPS Doppler time-series with a period of 47.8~days (Table \ref{tab:results2}) and an RV semi-amplitude of $\sim$1.8~\ms\ (Table \ref{tab:rv}) is caused by activity modulated by rotation. 
\subsection{System architecture and dynamical stability}
The ratio of the orbital period of TOI-763b and TOI-763c is 2.189, lies exterior to the 2:1 period commensurability. Planets in resonance are a sign that the planets migrated to their current observed location. Moreover, if the planets formed in the same location in a protoplanetary disc, it would be expected that they would have formed out of similar disc material and in this manner have comparable densities. Since the adjacent planets b and c have different densities this gives in addition hints that the planets formed in different locations in the protoplanetary disc and migrated inwards as e.g. in the Kepler-36 systems \citep{Carter+2012}.

The density of the outer serendipitously detected third planet candidate, d, close to a 4:1 period commensuability with \target c, is unknown since \tess\ did not observe a transit of this planet, and we therefore measured only a lower mass limit for it. If we define a transit as an event with impact parameter $\leq$ 1 (that is, we ignore very grazing transits), we get 89.05 degrees as an inclination limit for the outer planet to transit. 

We carried out a set of dynamical simulations in order to study the long-term stability of the system. Here we assumed all the signals to be of planetary origin, and we wanted to investigate if any of the parameters, in particular for the tentative planet "d" could be refined. The outer planet has no upper mass constraint and only a lower mass limit derived from the RVs of M$_\mathrm{d}$ $\sin~i$ = 9.54$\pm$1.59$M_\oplus$, assuming a circular orbit. We take the parameters reported in Table \ref{tab:results2} and draw 60,000 samples from the parameters posteriors as initial parameters for the dynamical simulation. Each parameter set was integrated for 10$^9$ orbits of the inner planet orbital period, covering the secular interaction timescale for the outer planet, using the Stability of the Planetary Orbital Configurations Klassifier (SPOCK) \citep{Tamayo2020a}. It was found that the system is dynamically stable for the whole parameter posterior space. The parameter space for the outer planet was studied in more detail whereby the true planetary mass is drawn from the reported $m \sin i$ in Table \ref{tab:results2} and allowing for inclination between 30 degrees,and 90 degrees, and with eccentricities up to 0.6. It is then found that stable systems can exist up to eccentricities 0.5 and for all tested inclination values. 
Therefore, additional observations are required to further confirm and constrain the parameters for the outer planet.
\section{Discussion}
\label{sec:disc}
 Data from \emph{Kepler} have shown that G- and K-stars tend to have at least one planet in an orbit with a period $<100$~days and that 
most such planets seem to be small. The most common types of planets so far tend to have masses of $\approx 2-10~M_\oplus$ and have radii with $R_P \approx 2-4~R_\oplus$ (mini-Neptunes) or with radii of $R_P \approx 1-2~R_\oplus$ (super-Earths), the latter thus having densities higher than those of the mini-Neptunes and potentially being rocky \citep{Marcy2014PNAS, Petigura+2017}.  
 When several planets are detected in such systems they are often found to be in  very compact arrangements where the ratio of two subsequent planet periods can often be below two.	

\target can be classified as a G-type star from its colors (Table~\ref{tab:stellar}). This is also confirmed by our spectral analysis in Section~\ref{sec:hoststar}. While generally, 
solar type stars are considered to be quite common, 
stars more massive than about G5 are rarer \citep{Adams2010}. Taken together with the fact that our own Sun belongs to this class of stars  makes the study of planet-hosting G-type objects quite worthwhile.  

The  planetary system accompanying \target~is, however, very 
different from our own. It consists of two confirmed planets with masses of $M_P(b)$ = 9.79 \mearth,  $M_P(c)$ =  9.32 \mearth and a possible planetary candidate consistent with a minimum mass of $M_P(d)/sin (i)$ = 9.54 \mearth in a very compact configuration, and with almost circular orbits with $a_b = 0.06$~AU, $a_c = 0.1$~AU, and  $a_d = 0.25$~AU corresponding to orbital 
periods of 5.6~days, 12.3~days and 47.8~days. The densities of the b and c planets are then 4.51 and 2.82 \gcc, respectively (Table~\ref{tab:results2}). 
This means that even the outermost planet in the TOI-763 system would be inside the orbit of the innermost planet, Mercury, in our own planet system. 
The orbital period ratio of planet b and c  is 2.2 which is  similar to the \emph{Kepler} compact systems, whereas the
orbital period ratio  of planet c and d is almost twice as high.
The TOI-763~planets are also much more massive than the innermost (rocky) planets in our solar system. With masses between 
9~\mearth and $10~M_\oplus$, they could all qualify as super-Earths with respect to mass. Being larger than twice the Earth's radii, both planets b and c have densities that could classify them as gaseous mini-Neptunes. This demonstrates that both accurate mass and radius are crucial in order to classify planets correctly, and ultimately will improve only with asteroseismology carried out from space \citep{Garcia+Ballot2019}. Since planet d has only a lower mass limit, it can either be a gaseous mini-Neptune or a rocky super Earth.

\begin{figure*}
	\includegraphics[width=0.8\textwidth]{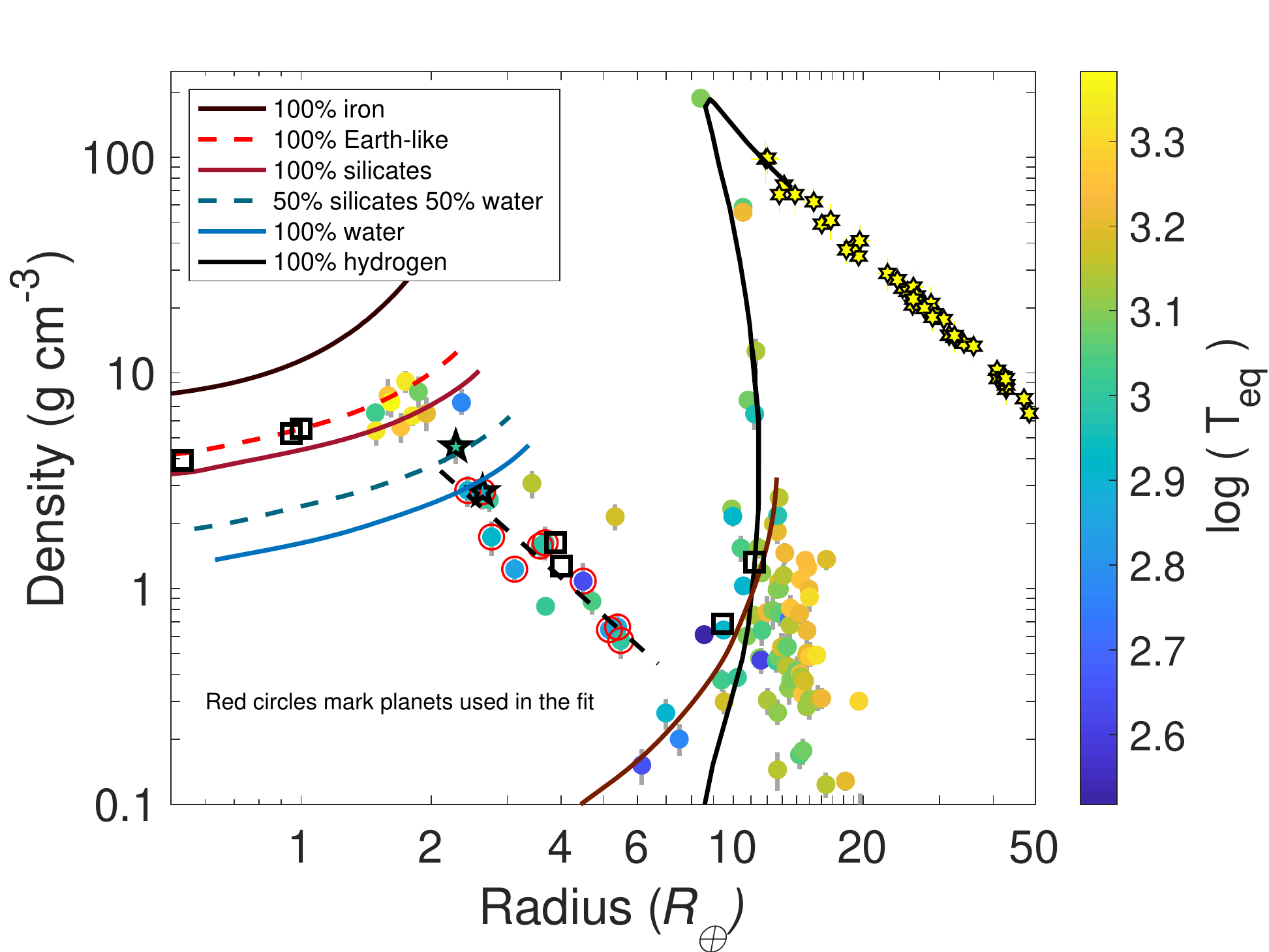}
        \caption{Density--radius diagram of planets orbiting G-type host stars with masses determined with at 
        least 15\% and radii with better than 5\% accuracy. The two star symbols represent \target b and c. The 
        black squares are the  solar system planets, and the bright yellow star symbols at radii between 12 and 50 \rearth~are 
      red dwarf stars from \citet{Persson2018}. The   theoretical mass-radius curves   are   from   \citet{2016ApJ...819..127Z}
       except the H-He model taken from \citet{Baraffe+2003, Baraffe+2008}.  
       The black dashed line  represent    a linear fit to the  ice planets marked with red circles (Sect.\ref{sec:disc}).
    \label{fig:Type42}
    }
\end{figure*}

In Fig.~\ref{fig:Type42} we plot a density-radius diagram of exoplanets orbiting solar type stars, here defined as having \teff~between 5300~K and 6000~K.  Since G-stars as host stars of exoplanets tend not to be selected for dedicated studies we choose to include only G-type stars in order to determine if they conform to the general patterns emerging in exoplanetology. To be able to determine trends in the diagram as precisely as the current data allows, and to have the same impact on density, we chose planets with  a precision in mass and radius better than 15~\% and 5~\%, respectively. We follow \cite{Persson2019} and \citet{Hatzes+Rauer2015} and include transiting brown dwarfs with measured masses \citep[cf. Table~6 in ][]{2020arXiv200201943C} in Fig.~\ref{fig:Type42}. We also include well determined eclipsing M-dwarfs \citep{Persson2019} which sets the upper mass of exoplanets. The planets are colour-coded with the logarithm of the planet equilibrium temperature assuming an albedo of zero. The planet data are downloaded from the NASA  exoplanet archive. We follow \citet{2020A&A...634A..43O} when assessing the best available measured parameters in the archive. We plot theoretical density-radius curves from iron to hydrogen \citep{2016ApJ...819..127Z}, and the H-He model from \citet{Baraffe+2003, Baraffe+2008}. 
The two star symbols mark the locations of TOI-763~b and c. We note that the planets b and c lie on opposite sides of the theoretical water ice line in Fig.~\ref{fig:Type42}, with c having the lower density. The solar system planets (except Mercury) are also included as black squares.

As immediately visible in  Fig.~\ref{fig:Type42}, planets fall into three separate areas of the diagram. Rocky planets 
can be seen in the upper left falling nicely close to the theoretical models. The ice-planets distribution follows closely a straight line from the rocky planets to the gas giants in the lower right where a turnoff is visible at a density of approximately 0.3~\gcc. The  gas giants, including the brown dwarfs, follow an almost vertical branch towards higher densities with almost a constant radii. This is caused by the interior electron degeneracy pressure which increase with mass until the point when the mass is sufficient to ignite hydrogen burning, causing a sharp turnoff at densities of approximately 200 \gcc (corresponding to $75-80$~\mjup). 
 As noted  
 in \citet{Persson2019}, the  gas giants with high $T_\mathrm{eq}$ have larger radii and thus lower densities 
 than predicted by models and falls off towards the lower right.

We fit a linear polynomial to the ice planets marked with red circles located between the theoretical water and hydrogen model lines 
and with an equilibrium temperature of  $T_\mathrm{eq} < 1000$~K.   
We find this polynomial to be described by \mbox{$\log \rho = -1.775\times \log (R) + 1.118$}. 
   This line would intersect the model tracks of the region between the 
  100~\% H and H-He models  at densities  of $0.2 -0.3$~g~cm$^{-3}$ 
  and radii of $8-10$~\rearth. This corresponds to a planet the size of Saturn, but with a density lower by more than a factor 
of two as clearly seen in Fig.~\ref{fig:Type42}. It is 
  especially interesting to find that even though Uranus and Neptune, as well as \target b, are excluded from the fit, they     
  fall almost perfectly on the  ``ice-track''. This  suggests that the solar system icy planets are similar to those found 
  outside of the solar system as far as  bulk density is concerned. What is interesting in terms of diversity is also 
  that almost all exoplanets   along the ``ice-track''   in our figure, have orbital periods less than 23~days (one
  planet, Kepler-396~c, has an orbital period of 88.5~days). 
This is in stark contrast to 
  Uranus and Neptune with
  orbital periods of 84~years and 165~years, respectively.

Having a mass roughly a factor of 1.5-2 lower than Neptune and Uranus, \target~b and c may have a different internal structure, since they are found on opposite sides of the water model in Fig.~\ref{fig:Type42}. Based on the two-layer silicate and water models of \citet{Zeng+Sasselov2014} and the \citet{2016ApJ...819..127Z} models plotted in Fig.~\ref{fig:Type42}, we can estimate that planet~b consists of a minimum of  $\sim40$~\% water and a maximum of $\sim60$~\% silicates, while planet~c has a density lower than the 100~\%  water models at the corresponding location in the diagram. This is, however, neglecting the possible existence of a thick gaseous H-He atmosphere that would thus increase the radius and lower the bulk density. Both planets could then have significant rocky cores surrounded by a gaseous envelope, the diameter of which, would depend on the history of energetic radiation and their respective location.
 These differences between planets b and c may indicate their formation in parts of the protostellar disk containing different types of material. The differences noted here, would clearly benefit from the possible carrying out of transmission spectroscopy of objects like \target b and c, using the JWST/NIRISS \citep{Doyon+2012}.
\section{Conclusions}

\label{sec:conc}
In this paper we have confirmed the planets, \target~b and c, found in the \tess light curves to be transiting \target, and we have further been able, through extensive radial velocity measurements,  to characterize them in terms of mass and radius. We find that both \target~b and c should contain large amounts of water, but demonstrate significant differences between them. We have also discovered a radial velocity signal that could be interpreted as one additional planet, which we tentatively indicate as \target~d. If confirmed by later work, we have found that it should have a similar (minimum) mass as planet~b and c, and an  orbital period  of 47.8~days. Planet~d is not detected in the \tess~photometric data.

Utilising the high-quality data for the planetary parameters, we have compared the data for planet~b and c, where we have determined high precision bulk densities, with nine other planets with equally high precision and that are also orbiting G-type main-sequence stars. We find that these planets all belong in the density regime of ``ice-planets'' and that their density versus radii distribution can be described by a first-degree polynomial with a very small scatter. All the planets, including \target b and c, that fall along the ``ice-track''
~and orbiting stars similar to our Sun, are found in the compact arrangement with short orbital periods, similar to what has been discovered so far for smaller planets orbiting low-mass stars in general.
\section{Data Availability}
The data underlying this article are available in the article and in its online supplementary material.

\section*{Acknowledgements} 
This work is done under the framework of the KESPRINT collaboration 
(http://kesprint.science). KESPRINT is an international consortium devoted to the characterisation and research of exoplanets discovered with space-based missions. M.F., C.M.P, and I.G. gratefully acknowledge the support of the  Swedish National Space Agency (DNR 65/19, 174/18). This work was supported by JSPS KAKENHI Grant Number JP20K14518. This is University of Texas Center for Planetary
Systems Habitability Contribution 0014. This work has made use of the VALD database, operated at Uppsala University, the Institute of Astronomy RAS in Moscow, and the University of Vienna. We are also indebted to N. Piskunov, and J. Valenti for the continued development and support of the Spectroscopy Made Easy (SME) package. This research has made use of the NASA Exoplanet Archive, which is operated by the California Institute of Technology, under contract with the National Aeronautics and Space Administration under the Exoplanet Exploration Program. This work is partly supported by JSPS KAKENHI Grant Numbers JP18H01265 and JP18H05439, and JST PRESTO Grant Number JPMJPR1775. L.M.S. and D.G. gratefully acknowledge financial support from the CRT foundation under Grant No. 2018.2323 ``Gaseous or rocky? Unveiling the nature of small worlds". K.W.F.L., J.K., Sz.Cs., M.E., S.G., A.P.H., M.P. and H.R. acknowledge support by DFG grants PA525/ 18-1, PA525/ 19-1, PA525/ 20-1, HA3279/ 12-1 and RA714/ 14-1 within the DFG Schwerpunkt SPP 1992, Exploring the Diversity of Extrasolar Planets. Resources supporting this work were provided by the NASA High-End Computing (HEC) Program through the NASA Advanced Supercomputing (NAS) Division at Ames Research Center for the production of the SPOC data products. P.K. and J.S. acknowledge the MSMT INTER-TRANSFER grant LTT20015. H.D. acknowledges support by grant ESP2017-87676-C5-4-R of the Spanish Secretary of State for R\&D\&i (MINECO).
Further, we also acknowledge the comment of an anonymous referee which improved the paper.



\bibliographystyle{mnras}
\bibliography{biblio} 



\section{List of affiliations}

$^{1}$Leiden Observatory, Leiden University, 2333CA Leiden, The Netherlands\\
$^{2}$Department of Space, Earth and Environment, \\
Chalmers University of Technology, Onsala Space Observatory, 439 92 Onsala, Sweden\\
$^{3}$Department of Astronomy, University of Tokyo, 7-3-1 Hongo, Bunkyo-ky, Tokyo 113-0033, Japan\\
$^{4}$Dipartimento di Fisica, Universitate degli Studi di Torino, via Pietro Giuria 1, I-10125, Torino, Italy\\
$^{5}$Center for Astronomy and Astrophysics, Technical University Berlin, Hardenbergstr. 36, 10623 Berlin, Germany\\
$^{6}$Vanderbilt University, Physics and Astronomy Department, Nashville, TN 37235, USA\\
$^{7}$Astrophysics Group, Keele University, Staffordshire, ST5 5BG, UK\\
$^{8}$Rheinisches Institut f\"ur Umweltforschung an der Universit\"at zu K\"oln, Aachener Strasse 209, D-50931 K\"oln, Germany\\
$^{9}$Th\"uringer Landessternwarte Tautenburg, 07778, Tautenburg, Germany\\
$^{10}$ Dipartimento di Fisica e Astronomia Galilei, Universit\'a di Padova, Vicolo dell'Osservatorio 3, I-35122 Padova, Italy \\
$^{11}$Instituto de Astrofisica de Canarias, C/ Via Lactea s/n, E-38205 La Laguna, Spain\\
$^{12}$Departamento de Astrofisica, Universidad de La Laguna, E-38206 La Laguna, Spain\\
$^{13}$Astronomy Department and Van Vleck Observatory, Wesleyan University, Middletown, CT 06459, USA\\
$^{14}$Stellar Astrophysics Centre, Department of Physics and Astronomy, Aarhus University, Ny Munkegade 120, DK-8000 Aarhus C, Denmark\\
$^{15}$Sub-department of Astrophysics, Department of Physics, University of Oxford, Oxford OX1 3RH, UK\\
$^{16}$INAF - Osservatorio Astronomico di Palermo, Piazza del Parlamento, 1, I-90134 Palermo, Italy\\
$^{17}$Department of Astrophysical Sciences, Princeton University, 4 Ivy Lane, Princeton, NJ 08544, USA\\
$^{18}$Institute of Planetary Research, German Aerospace Center, Rutherfordstrasse 2, D-12489 Berlin, Germany\\
$^{19}$Department of Astronomy and McDonald Observatory, University of Texas at Austin, 2515 Speedway, Stop C1400, Austin, TX 78712, USA\\
$^{20}$Center for Planetary Systems Habitability, University of Texas at Austin, Austin, TX 78712, USA\\
$^{21}$Department of Physics and Kavli Institute for Astrophysics and Space Research, Massachusetts Institute of Technology, Cambridge, MA 02139, USA\\
$^{22}$Department of Earth and Planetary Sciences, Tokyo Institute of Technology, 2-12-1 Ookayama, Meguro-ku, Tokio 152-8551, Japan\\
$^{23}$NASA Ames Research Center, Moffet Field, CA 94035, USA\\
$^{24}$Astronomical Institute, Czech Academy of Sciences, Fri\v{c}ova 298, 25165, Ond\v{r}ejov, Czech Republic\\
$^{25}$Center for Astrophysics | Harvard \& Smithsonian, 60 Garden Street, Cambridge, MA 02138, USA
$^{26}$National Astronomical Observatory of Japan, NINS, 2-21-1 Osawa, Mitaka, Tokyo 1818588, Japan\\
$^{27}$Komaba Institute for Science, The University of Tokyo, 3-8-1 Komaba, Meguro, Tokyo 153-8902, Japan\\
$^{28}$JST, PRESTO, 3-8-1 Komaba, Meguro, Tokyo 153-8902, Japan\\
$^{29}$Astrobiology Center, NINS, 2-21-1 Osawa, Mitaka, Tokyo 181-8588, Japan\\
$^{30}$European Southern Observatory (ESO), Alonso de C\'ordova 3107, Vitacura, Casilla 19001, Santiago de Chile, Chile\\
$^{31}$Department of Earth, Atmospheric and Planetary Sciences, Massachusetts Institute of Technology, Cambridge, MA 02139, USA\\
$^{32}$Department of Aeronautics and Astronautics, MIT, 77 Massachusetts Avenue, Cambridge, MA 02139, USA\\
$^{33}$School of Physics and Astronomy, Monash University, VIC 3800, Australia and ARC Centre of Excellence for All Sky Astrophysics in Three Dimensions (ASTRO-3D)\\
$^{34}$Astronomical Institute, Faculty of Mathematics and Physics, Charles University, Ke Karlovu 2027/3, 12116 Prague, Czech Republic\\
$^{35}$SETI Institute/NASA Ames Research Center, Moffet Field, CA 94035, USA\\
$^{36}$Mullard Space Science Laboratory, University College London, Holmbury St Mary, Dorking, Surrey RH5 6NT, UK\\
$^{37}$Space Telescope Science Institute, Baltimore, MD 21218, USA\\
$^{38}$Institut fuer Geologische Wissenschaften, Freie Universitaet Berlin, 12249 Berlin, Germany\\
$^{39}$Department of Astronomy, University of Wisconsin-Madison, Madison, WI 53706, USA

\appendix

\section{The radial velocity data}

\begin{table*} 
\caption{\target's HARPS RVs, CCF's bisector inverse slope (BIS), CCF's full-width at half maximum (FWHM), CCF's contrast, and Ca {\sc ii} H \& K line activity indicator (\logrhk) as extracted using the HARPS data reduction software (\texttt{DRS}). The exposure time and S/N ratio per pixel at 5500\,\AA\ are listed in the last two columns.}
\label{table:DRS_RVs}
\centering
\scriptsize
\begin{tabular}{cccccccccc}
\hline\hline
BJD$_\mathrm{TDB}$ &    RV  &   $\sigma_\mathrm{RV}$   &  BIS   &  FWHM  & Contrast & log\,R$^\prime_\mathrm{HK}$ & $\sigma_\mathrm{log\,R^\prime_\mathrm{HK}}$ & T$_\mathrm{exp}$ & SNR per pix. \\
$-2450000$ (d)     & (\kms) & (\kms) & (\kms) & (\kms) &     &     &      &       (s)        &  @5550\,\AA    \\
\hline
 8655.574286 & -13.9795 & 0.0019 & -0.0277 & 6.8118 & 49.046 & -5.005 & 0.025 & 1800 &  47.5 \\
 8657.585518 & -13.9857 & 0.0012 & -0.0216 & 6.8193 & 48.978 & -4.964 & 0.014 & 1800 &  74.5 \\
 8660.605351 & -13.9818 & 0.0018 & -0.0288 & 6.8142 & 49.065 & -5.020 & 0.030 & 1800 &  52.8 \\
 8661.499253 & -13.9813 & 0.0012 & -0.0322 & 6.8216 & 49.053 & -4.958 & 0.014 & 1800 &  73.8 \\
 8661.631697 & -13.9812 & 0.0015 & -0.0293 & 6.8157 & 49.056 & -5.048 & 0.028 & 1800 &  63.5 \\
 8664.593171 & -13.9844 & 0.0011 & -0.0228 & 6.8287 & 49.027 & -4.960 & 0.015 & 1800 &  85.4 \\
 8666.546629 & -13.9784 & 0.0013 & -0.0310 & 6.8271 & 49.038 & -4.976 & 0.018 & 1800 &  70.0 \\
 8667.533554 & -13.9784 & 0.0012 & -0.0252 & 6.8165 & 49.039 & -4.955 & 0.017 & 1800 &  76.9 \\
 8667.608594 & -13.9799 & 0.0011 & -0.0252 & 6.8171 & 49.056 & -4.966 & 0.016 & 1800 &  81.0 \\
 8668.562153 & -13.9783 & 0.0017 & -0.0231 & 6.8288 & 48.991 & -4.965 & 0.023 & 1800 &  52.7 \\
 8669.518396 & -13.9798 & 0.0011 & -0.0212 & 6.8221 & 48.966 & -4.920 & 0.009 & 1800 &  78.3 \\
 8669.594177 & -13.9798 & 0.0011 & -0.0288 & 6.8187 & 48.947 & -4.934 & 0.011 & 1800 &  81.7 \\
 8670.536138 & -13.9782 & 0.0012 & -0.0267 & 6.8167 & 49.020 & -4.951 & 0.015 & 1800 &  76.5 \\
 8670.606619 & -13.9786 & 0.0013 & -0.0319 & 6.8250 & 48.997 & -4.982 & 0.017 & 1800 &  72.0 \\
 8673.578403 & -13.9841 & 0.0014 & -0.0268 & 6.8163 & 48.964 & -4.998 & 0.020 & 1800 &  64.9 \\
 8674.584085 & -13.9872 & 0.0014 & -0.0249 & 6.8324 & 48.908 & -4.989 & 0.021 & 1800 &  65.9 \\
 8676.495617 & -13.9777 & 0.0014 & -0.0261 & 6.8243 & 48.996 & -4.984 & 0.020 & 1800 &  66.3 \\
 8676.567243 & -13.9795 & 0.0015 & -0.0356 & 6.8266 & 48.967 & -4.985 & 0.025 & 1800 &  62.7 \\
 8677.509955 & -13.9752 & 0.0012 & -0.0314 & 6.8231 & 49.009 & -5.037 & 0.020 & 1800 &  75.0 \\
 8678.584762 & -13.9797 & 0.0016 & -0.0351 & 6.8174 & 49.041 & -5.018 & 0.030 & 1800 &  59.4 \\
 8679.531662 & -13.9801 & 0.0010 & -0.0285 & 6.8210 & 49.006 & -5.005 & 0.015 & 1800 &  91.0 \\
 8679.596749 & -13.9812 & 0.0020 & -0.0403 & 6.8162 & 49.030 & -5.072 & 0.055 & 1800 &  50.8 \\
 8680.539773 & -13.9784 & 0.0008 & -0.0282 & 6.8238 & 49.016 & -4.981 & 0.011 & 1800 & 108.4 \\
 8681.547466 & -13.9777 & 0.0014 & -0.0257 & 6.8198 & 49.005 & -5.029 & 0.024 & 1800 &  66.0 \\
 8682.528206 & -13.9745 & 0.0010 & -0.0263 & 6.8223 & 49.020 & -4.982 & 0.013 & 1800 &  86.5 \\
 8682.576813 & -13.9726 & 0.0011 & -0.0238 & 6.8221 & 49.004 & -4.983 & 0.016 & 1800 &  80.2 \\
 8683.470732 & -13.9776 & 0.0013 & -0.0306 & 6.8303 & 48.968 & -4.981 & 0.017 & 1800 &  70.5 \\
 8683.490255 & -13.9793 & 0.0013 & -0.0309 & 6.8131 & 49.034 & -4.957 & 0.016 & 1800 &  69.1 \\
 8684.558927 & -13.9866 & 0.0012 & -0.0280 & 6.8276 & 48.949 & -4.972 & 0.015 & 1800 &  73.7 \\
 8684.621364 & -13.9833 & 0.0014 & -0.0340 & 6.8268 & 48.995 & -5.020 & 0.024 & 1800 &  67.7 \\
 8685.519032 & -13.9856 & 0.0012 & -0.0276 & 6.8247 & 49.023 & -4.983 & 0.018 & 1800 &  76.4 \\
 8685.565024 & -13.9854 & 0.0033 & -0.0305 & 6.8062 & 49.075 & -5.141 & 0.114 & 1800 &  34.2 \\
 8689.528163 & -13.9813 & 0.0015 & -0.0230 & 6.8234 & 49.019 & -4.978 & 0.022 & 1800 &  64.0 \\
 8689.572326 & -13.9783 & 0.0017 & -0.0301 & 6.8198 & 49.038 & -5.053 & 0.036 & 1800 &  55.6 \\
 8690.507814 & -13.9822 & 0.0015 & -0.0319 & 6.8228 & 49.016 & -5.009 & 0.022 & 1800 &  60.5 \\
 8690.532153 & -13.9843 & 0.0013 & -0.0278 & 6.8234 & 49.061 & -4.958 & 0.017 & 1800 &  67.0 \\
 8691.510994 & -13.9815 & 0.0012 & -0.0298 & 6.8184 & 49.040 & -5.001 & 0.018 & 2100 &  76.5 \\
 8692.539322 & -13.9819 & 0.0016 & -0.0299 & 6.8223 & 49.070 & -5.013 & 0.024 & 1800 &  57.9 \\
 8693.488271 & -13.9758 & 0.0013 & -0.0309 & 6.8227 & 49.064 & -4.991 & 0.018 & 1800 &  71.7 \\
 8694.515083 & -13.9774 & 0.0012 & -0.0274 & 6.8182 & 49.047 & -4.977 & 0.016 & 1500 &  74.7 \\
 8694.532628 & -13.9775 & 0.0012 & -0.0240 & 6.8225 & 49.040 & -4.986 & 0.016 & 1500 &  79.9 \\
 8695.485141 & -13.9845 & 0.0011 & -0.0304 & 6.8261 & 49.033 & -4.963 & 0.013 & 1800 &  84.0 \\
 8697.505285 & -13.9883 & 0.0018 & -0.0292 & 6.8194 & 49.012 & -4.967 & 0.029 & 1800 &  53.7 \\
 8697.526684 & -13.9906 & 0.0019 & -0.0298 & 6.8219 & 49.028 & -5.043 & 0.038 & 1800 &  51.6 \\
 8698.514889 & -13.9834 & 0.0012 & -0.0290 & 6.8191 & 49.020 & -4.967 & 0.016 & 1800 &  73.4 \\
 8698.534830 & -13.9875 & 0.0013 & -0.0271 & 6.8231 & 49.023 & -4.942 & 0.016 & 1800 &  72.8 \\
 8699.502250 & -13.9840 & 0.0012 & -0.0251 & 6.8228 & 49.024 & -4.978 & 0.017 & 1800 &  79.1 \\
 8700.490814 & -13.9815 & 0.0012 & -0.0245 & 6.8143 & 49.002 & -4.999 & 0.017 & 1800 &  76.6 \\
 8700.511796 & -13.9830 & 0.0012 & -0.0196 & 6.8163 & 49.002 & -4.954 & 0.015 & 1800 &  79.5 \\
 8701.515464 & -13.9844 & 0.0011 & -0.0290 & 6.8184 & 48.995 & -4.962 & 0.015 & 1800 &  89.4 \\
 8702.518461 & -13.9864 & 0.0014 & -0.0270 & 6.8170 & 48.974 & -5.002 & 0.024 & 1800 &  66.4 \\
 8703.523842 & -13.9836 & 0.0015 & -0.0332 & 6.8290 & 48.929 & -4.988 & 0.025 & 2100 &  61.9 \\
 8704.499781 & -13.9773 & 0.0020 & -0.0286 & 6.8309 & 48.910 & -5.029 & 0.038 & 2400 &  49.3 \\
 8705.474923 & -13.9761 & 0.0010 & -0.0282 & 6.8175 & 49.008 & -4.991 & 0.014 & 1800 &  93.7 \\
 8706.476486 & -13.9814 & 0.0011 & -0.0266 & 6.8184 & 48.972 & -4.965 & 0.015 & 1800 &  82.4 \\
 8707.484588 & -13.9868 & 0.0017 & -0.0295 & 6.8184 & 48.944 & -4.996 & 0.026 & 2400 &  56.5 \\
 8708.490874 & -13.9869 & 0.0012 & -0.0341 & 6.8251 & 48.960 & -4.986 & 0.017 & 1800 &  77.7 \\
 8709.483806 & -13.9848 & 0.0016 & -0.0306 & 6.8147 & 49.000 & -4.975 & 0.026 & 1800 &  60.8 \\
 8710.481089 & -13.9787 & 0.0015 & -0.0305 & 6.8206 & 49.023 & -5.008 & 0.024 & 1800 &  63.8 \\
 8710.499999 & -13.9843 & 0.0018 & -0.0287 & 6.8191 & 48.971 & -5.023 & 0.034 & 1800 &  53.5 \\
 8711.475294 & -13.9794 & 0.0013 & -0.0270 & 6.8222 & 48.954 & -4.979 & 0.018 & 2100 &  75.0 \\
 8712.492704 & -13.9813 & 0.0018 & -0.0269 & 6.8286 & 48.936 & -5.004 & 0.030 & 2100 &  54.6 \\
 8713.481667 & -13.9840 & 0.0014 & -0.0276 & 6.8160 & 48.985 & -4.956 & 0.019 & 2100 &  68.0 \\
 8715.477396 & -13.9778 & 0.0012 & -0.0292 & 6.8233 & 48.968 & -4.945 & 0.014 & 2400 &  79.3 \\
 8716.475018 & -13.9724 & 0.0010 & -0.0271 & 6.8255 & 48.960 & -4.933 & 0.012 & 1800 &  92.8 \\
 8717.475742 & -13.9776 & 0.0018 & -0.0282 & 6.8276 & 48.908 & -4.968 & 0.028 & 1800 &  52.0 \\
 8718.474720 & -13.9834 & 0.0012 & -0.0229 & 6.8220 & 48.959 & -4.933 & 0.015 & 1800 &  77.1 \\
 8719.476638 & -13.9854 & 0.0013 & -0.0261 & 6.8197 & 48.954 & -5.006 & 0.021 & 1800 &  74.3 \\
 8721.475163 & -13.9764 & 0.0013 & -0.0260 & 6.8147 & 48.967 & -4.968 & 0.020 & 1800 &  72.0 \\
 8722.480011 & -13.9826 & 0.0011 & -0.0296 & 6.8186 & 49.002 & -4.976 & 0.017 & 1800 &  84.1 \\
 8724.475162 & -13.9843 & 0.0009 & -0.0288 & 6.8192 & 48.988 & -4.957 & 0.010 & 1800 & 100.6 \\
 8725.476170 & -13.9851 & 0.0011 & -0.0250 & 6.8168 & 49.007 & -4.970 & 0.016 & 1800 &  86.9 \\
 8727.475528 & -13.9741 & 0.0017 & -0.0340 & 6.8129 & 48.956 & -5.097 & 0.036 & 1800 &  58.6 \\
 8728.477649 & -13.9758 & 0.0015 & -0.0330 & 6.8221 & 48.960 & -5.013 & 0.025 & 2100 &  62.3 \\
\hline
\end{tabular}
\end{table*}

\begin{table*}
\caption{\target's HARPS differential line width (dLW), chromatic index (Crx), and Na D and H$\alpha$ line activity indicators extracted with the spectrum radial velocity analyser (\texttt{SERVAL}).}
\label{table:Serval_Indexes}
\centering
\scriptsize
\begin{tabular}{crrrrrrrr}
\hline\hline
BJD$_\mathrm{TDB}$ &    dLW~~\  &   $\sigma_\mathrm{dLW}$~~\   &  Crx~~~~~~\   &  $\sigma_\mathrm{Crx}$~~~~~~\  & H$\alpha$~~~\ & $\sigma_\mathrm{H\alpha}$~~\ & Na D~\ & $\sigma_\mathrm{Na\,D}$  \\
$-2450000$ (d)& (m$^2$\,s$^{-2}$) & (m$^2$\,s$^{-2}$) & (m~s$^{-1}$\,Np$^{-1}$) & (m~s$^{-1}$\,Np$^{-1}$)&            &                                   &      &                       \\
\hline
 8655.574286 &   -27.3234 &     3.4815 &   -13.6056 &    13.3345 &     0.4439 &     0.0018 &     0.2781 &     0.0021 \\
 8657.585518 &   -34.9508 &     2.2257 &    -9.8812 &    11.1881 &     0.4416 &     0.0011 &     0.2719 &     0.0012 \\
 8660.605351 &   -34.1636 &     3.0802 &   -13.6919 &    13.9300 &     0.4436 &     0.0016 &     0.2781 &     0.0018 \\
 8661.499253 &   -30.3810 &     2.4287 &    -4.0789 &    10.7817 &     0.4395 &     0.0011 &     0.2774 &     0.0012 \\
 8661.631697 &   -32.4117 &     2.7012 &    17.6440 &    14.2956 &     0.4429 &     0.0013 &     0.2735 &     0.0014 \\
 8664.593171 &   -30.4048 &     1.9665 &   -10.5227 &     8.8589 &     0.4398 &     0.0009 &     0.2799 &     0.0011 \\
 8666.546629 &   -29.9505 &     1.9517 &    11.0804 &    11.7434 &     0.4410 &     0.0012 &     0.2726 &     0.0013 \\
 8667.533554 &   -30.8120 &     2.2702 &     8.4083 &    11.1142 &     0.4408 &     0.0010 &     0.2761 &     0.0012 \\
 8667.608594 &   -33.0404 &     2.3743 &     1.7582 &    10.8517 &     0.4383 &     0.0010 &     0.2727 &     0.0011 \\
 8668.562153 &   -27.1619 &     3.3684 &    -2.5544 &    14.7873 &     0.4438 &     0.0016 &     0.2765 &     0.0018 \\
 8669.518396 &   -19.2356 &     2.1273 &    -3.4056 &    10.8156 &     0.4423 &     0.0011 &     0.2752 &     0.0012 \\
 8669.594177 &   -19.1052 &     2.0280 &    -3.7332 &    11.8615 &     0.4406 &     0.0010 &     0.2813 &     0.0011 \\
 8670.536138 &   -29.0092 &     2.2689 &    -2.7832 &    10.5860 &     0.4414 &     0.0010 &     0.2801 &     0.0012 \\
 8670.606619 &   -24.9930 &     2.1656 &    -0.1078 &    12.1134 &     0.4394 &     0.0011 &     0.2785 &     0.0013 \\
 8673.578403 &   -26.6192 &     2.3838 &     2.4626 &    12.9057 &     0.4382 &     0.0013 &     0.2756 &     0.0014 \\
 8674.584085 &   -24.7986 &     2.9662 &     6.7085 &    11.4400 &     0.4401 &     0.0012 &     0.2717 &     0.0014 \\
 8676.495617 &   -28.5964 &     2.6941 &   -12.1365 &    11.3112 &     0.4375 &     0.0012 &     0.2715 &     0.0014 \\
 8676.567243 &   -26.4019 &     3.5562 &    -8.3655 &    13.7688 &     0.4384 &     0.0013 &     0.2728 &     0.0015 \\
 8677.509955 &   -22.6246 &     2.7620 &    -5.5261 &     9.9081 &     0.4424 &     0.0011 &     0.2747 &     0.0012 \\
 8678.584762 &   -17.4065 &     3.2822 &     0.0232 &    13.4401 &     0.4430 &     0.0013 &     0.2810 &     0.0016 \\
 8679.531662 &   -29.2413 &     2.3215 &    -4.0168 &     9.9460 &     0.4418 &     0.0009 &     0.2770 &     0.0010 \\
 8679.596749 &   -15.9057 &     4.5005 &    33.3329 &    16.8297 &     0.4435 &     0.0015 &     0.2836 &     0.0018 \\
 8680.539773 &   -26.7658 &     1.9028 &   -10.6048 &     9.3233 &     0.4404 &     0.0007 &     0.2767 &     0.0008 \\
 8681.547466 &   -18.2137 &     3.4352 &     7.4320 &    11.0308 &     0.4402 &     0.0012 &     0.2787 &     0.0014 \\
 8682.528206 &   -24.8256 &     2.1641 &    -0.9159 &     7.0255 &     0.4388 &     0.0009 &     0.2775 &     0.0010 \\
 8682.576813 &   -28.1810 &     2.1110 &     1.8850 &    10.6389 &     0.4392 &     0.0010 &     0.2767 &     0.0011 \\
 8683.470732 &   -25.2732 &     2.7230 &    -4.5382 &    11.2486 &     0.4413 &     0.0011 &     0.2799 &     0.0013 \\
 8683.490255 &   -28.7430 &     2.5488 &    -2.2769 &    13.2397 &     0.4414 &     0.0012 &     0.2791 &     0.0013 \\
 8684.558927 &   -24.7389 &     2.5555 &    -5.2513 &    11.5722 &     0.4427 &     0.0011 &     0.2731 &     0.0012 \\
 8684.621364 &   -24.3193 &     3.1360 &    -0.9628 &    10.2176 &     0.4422 &     0.0012 &     0.2702 &     0.0013 \\
 8685.519032 &   -29.4160 &     2.6435 &     8.1137 &     9.9926 &     0.4414 &     0.0010 &     0.2791 &     0.0012 \\
 8685.565024 &   -32.8843 &     6.6776 &    26.1068 &    25.8119 &     0.4436 &     0.0024 &     0.2793 &     0.0029 \\
 8689.528163 &   -27.6250 &     2.3541 &     3.5240 &    16.1837 &     0.4412 &     0.0012 &     0.2746 &     0.0014 \\
 8689.572326 &   -32.6538 &     3.2969 &    -3.0382 &    20.6879 &     0.4400 &     0.0014 &     0.2744 &     0.0017 \\
 8690.507814 &   -31.3709 &     2.8905 &     5.1231 &    13.7563 &     0.4423 &     0.0014 &     0.2755 &     0.0016 \\
 8690.532153 &   -32.3343 &     2.8776 &     8.9537 &    12.6298 &     0.4417 &     0.0012 &     0.2734 &     0.0014 \\
 8691.510994 &   -33.8941 &     2.6392 &   -12.2249 &    10.9119 &     0.4389 &     0.0010 &     0.2710 &     0.0012 \\
 8692.539322 &   -34.7286 &     3.1755 &     7.5525 &    12.9709 &     0.4382 &     0.0014 &     0.2763 &     0.0016 \\
 8693.488271 &   -32.6269 &     2.6853 &    10.5369 &    12.3633 &     0.4394 &     0.0011 &     0.2765 &     0.0013 \\
 8694.515083 &   -30.1416 &     2.7134 &     2.8462 &     9.0025 &     0.4388 &     0.0011 &     0.2784 &     0.0012 \\
 8694.532628 &   -29.5008 &     2.4497 &     0.0987 &     9.5962 &     0.4415 &     0.0010 &     0.2784 &     0.0011 \\
 8695.485141 &   -30.2703 &     2.2118 &    -2.0782 &     9.6373 &     0.4397 &     0.0010 &     0.2769 &     0.0011 \\
 8697.505285 &   -32.9234 &     3.7203 &    19.9173 &    11.3887 &     0.4393 &     0.0015 &     0.2829 &     0.0017 \\
 8697.526684 &   -32.4152 &     3.7858 &    14.7366 &    13.3673 &     0.4397 &     0.0015 &     0.2817 &     0.0018 \\
 8698.514889 &   -32.5441 &     2.5370 &    -5.6555 &    10.0501 &     0.4411 &     0.0011 &     0.2801 &     0.0012 \\
 8698.534830 &   -31.2862 &     2.6793 &    -0.8187 &     9.5557 &     0.4404 &     0.0011 &     0.2786 &     0.0012 \\
 8699.502250 &   -31.5803 &     2.3837 &    -7.1028 &    10.2506 &     0.4383 &     0.0010 &     0.2738 &     0.0011 \\
 8700.490814 &   -30.0590 &     2.4068 &     0.5601 &     9.8587 &     0.4404 &     0.0010 &     0.2729 &     0.0012 \\
 8700.511796 &   -32.0105 &     2.0274 &     9.6646 &     9.7858 &     0.4382 &     0.0010 &     0.2745 &     0.0011 \\
 8701.515464 &   -32.4156 &     2.0589 &     0.2134 &     7.9769 &     0.4405 &     0.0008 &     0.2706 &     0.0010 \\
 8702.518461 &   -32.8061 &     2.8615 &   -19.3266 &    12.4775 &     0.4389 &     0.0012 &     0.2723 &     0.0014 \\
 8703.523842 &   -26.9803 &     3.0420 &    -6.9396 &    13.1684 &     0.4396 &     0.0013 &     0.2718 &     0.0015 \\
 8704.499781 &   -24.2002 &     4.3921 &     3.4798 &    17.7909 &     0.4395 &     0.0016 &     0.2860 &     0.0019 \\
 8705.474923 &   -28.2223 &     1.7840 &    -1.9189 &     7.7832 &     0.4402 &     0.0008 &     0.2776 &     0.0010 \\
 8706.476486 &   -29.3501 &     2.2899 &     5.5727 &     8.8095 &     0.4388 &     0.0009 &     0.2710 &     0.0011 \\
 8707.484588 &   -15.5731 &     3.1948 &    -1.2679 &    14.0285 &     0.4388 &     0.0014 &     0.2744 &     0.0016 \\
 8708.490874 &   -23.6189 &     2.7607 &     1.1063 &     9.5419 &     0.4418 &     0.0010 &     0.2710 &     0.0012 \\
 8709.483806 &   -14.6217 &     3.6523 &    14.3927 &    12.0294 &     0.4420 &     0.0013 &     0.2736 &     0.0015 \\
 8710.481089 &   -23.0342 &     3.2025 &    29.8668 &    13.5686 &     0.4396 &     0.0012 &     0.2806 &     0.0014 \\
 8710.499999 &    -7.3168 &     4.7679 &    44.4273 &    13.5357 &     0.4399 &     0.0015 &     0.2775 &     0.0017 \\
 8711.475294 &   -16.1286 &     3.1201 &    -4.7268 &    11.7724 &     0.4403 &     0.0010 &     0.2843 &     0.0012 \\
 8712.492704 &   -18.6837 &     3.4692 &    -1.9183 &    14.8572 &     0.4413 &     0.0014 &     0.2772 &     0.0017 \\
 8713.481667 &   -27.0469 &     2.8001 &    -5.2989 &    11.5990 &     0.4411 &     0.0012 &     0.2762 &     0.0014 \\
 8715.477396 &   -22.7484 &     2.3661 &    -6.6190 &    10.1799 &     0.4419 &     0.0010 &     0.2795 &     0.0011 \\
 8716.475018 &   -25.0995 &     2.1692 &     2.4860 &     9.2656 &     0.4423 &     0.0008 &     0.2714 &     0.0010 \\
 8717.475742 &   -21.9609 &     3.8755 &    -1.4322 &    17.3181 &     0.4410 &     0.0015 &     0.2744 &     0.0018 \\
 8718.474720 &   -23.7491 &     2.3748 &     3.2041 &    11.6205 &     0.4408 &     0.0010 &     0.2727 &     0.0012 \\
 8719.476638 &   -29.0434 &     2.8562 &   -10.6933 &    12.1987 &     0.4420 &     0.0010 &     0.2795 &     0.0012 \\
 8721.475163 &   -27.9779 &     2.6364 &     2.3398 &    11.0052 &     0.4414 &     0.0011 &     0.2722 &     0.0012 \\
 8722.480011 &   -27.7482 &     2.3911 &    -8.0116 &    10.3813 &     0.4399 &     0.0009 &     0.2801 &     0.0011 \\
 8724.475162 &   -26.6539 &     2.1566 &     1.0507 &     7.3636 &     0.4401 &     0.0008 &     0.2721 &     0.0009 \\
 8725.476170 &   -25.7527 &     2.2573 &     5.1373 &     8.0819 &     0.4395 &     0.0009 &     0.2730 &     0.0010 \\
 8727.475528 &   -18.7502 &     3.8639 &    24.3561 &    14.6378 &     0.4404 &     0.0013 &     0.2798 &     0.0016 \\
 8728.477649 &   -24.0673 &     3.1167 &    20.2222 &    12.1625 &     0.4415 &     0.0013 &     0.2803 &     0.0015 \\
 \hline
\end{tabular}
\end{table*}


\bsp	
\label{lastpage}
\end{document}